\documentclass[journal,draftclsnofoot,onecolumn,12pt]{IEEEtran}
\usepackage{amsthm,amssymb,graphicx,multirow,amsmath,color,amsfonts}
\usepackage[update,prepend]{epstopdf}
\usepackage{cite}
\usepackage[latin1]{inputenc}
\usepackage{tikz}
\usepackage{bbm} 
\usepackage{pdfpages}
\usepackage{tabulary}
\usepackage{multirow}
\usepackage{comment}
\usepackage{mathtools}
\usepackage{dsfont}
\usepackage{caption}
\usepackage{subcaption}
\usepackage{eucal}
\usepackage{algorithm}
\usepackage{algpseudocode}
\usepackage{nccmath}
\usepackage{mathtools, cuted}\usepackage{lipsum}
\DeclareMathOperator\erf{erf}
\DeclareMathOperator\arcos{arcos}
\DeclareMathOperator\lognormal{lognormal}

\def\nb0{{\mathbf{0}}}
\def\nb1{{\mathbf{1}}}

\newtheorem{lemma}{Lemma}

\newtheorem{prop}{Proposition}

\newtheorem{remark}{Remark}

\def\erf{\operatorname{erf}}

\allowdisplaybreaks 

\allowdisplaybreaks 
\usepackage{setspace}	
\begin{document}
\graphicspath{{./Figures/}}
\title{
Laser-Powered UAVs for Wireless Communication Coverage: A Large-Scale Deployment Strategy}

\author{Mohamed-Amine~Lahmeri,~\IEEEmembership{Student Member,~IEEE,}
	Mustafa~A. Kishk,~\IEEEmembership{Member,~IEEE,}
	and~ Mohamed-Slim~Alouini,~\IEEEmembership{Fellow,~IEEE}
	\thanks{Mohamed-Amine Lahmeri was with King Abdullah University of Science and Technology (KAUST), Computer, Electrical, and Mathematical Science and Engineering (CEMSE) division, when he carried this work. Currently, he is a Ph.D. student at the Institute for Digital
		Communications, Friedrich-Alexander-Universit\"at
		Erlangen-N\"urnberg,
		Germany; Mustafa Kishk is with the Department of Electronic Engineering, National University of Ireland, Maynooth, W23 F2H6, Ireland. Mohamed-Slim Alouini is with the CEMSE Division, KAUST, Thuwal 23955, Saudi Arabia (e-mail: amine.lahmeri@fau.de;
		mustafa.kishk@mu.ie; slim.alouini@kaust.edu.sa).}
}

\maketitle
\begin{abstract}
The use of unmanned aerial vehicles (UAVs) is strongly advocated for sixth-generation (6G) networks, as the 6G standard will not be limited to improving broadband services, but will also target the extension of the geographical cellular coverage. In this context, the deployment of UAVs is considered a key solution for seamless connectivity and reliable coverage. That being said, it is important to underline that although UAVs are characterized by their high mobility and their ability to establish line-of-sight (LOS) links, their use is still impeded by several factors such as weather conditions, their limited computing power, and, most importantly, their limited energy. In this work, we are aiming for the novel technology that enables indefinite wireless power transfer for UAVs using laser beams. We propose a novel UAV deployment strategy, based on which we analyze the overall performance of the system in terms of wireless coverage. To this end, we use tractable tools from stochastic geometry to model the complex communication system. We analyze the user's connectivity profile under different laser charging capabilities and in different type of environments. We show a decrease in the coverage probability by more than 12\% in moderate-to-strong turbulence conditions compared to low turbulence conditions. We also show how the connection rate to the aerial network significantly decreases in favor of the terrestrial network for short laser charging ranges. We conclude that laser-powered drones are considered interesting alternatives when placed in LOS with users, in low-to-moderate optical turbulence, and at reasonable ranges from the charging stations.
\end{abstract}
 
\vspace{-4mm}
\section{Introduction} \label{sec:intro}
With the recent upsurge in the demand for fast and reliable connectivity, researchers have been gravitating incessantly towards integrating aerial and terrestrial architectures. It is expected that low latency, seamless connection, and wide coverage areas will be a real paramount to sixth-generation (6G) networks~\cite{AlouiniVision,SaadVision}. In this context, unmanned aerial vehicles (UAVs) are deemed to be the road to meet the aforementioned requirements of the future 6G standards in light of their multiple striking features. For instance, we mention their high flexibility and ability to establish line-of-sight (LOS) links with users. Without a doubt, drones were used in a plethora of applications such as in the military, farming, traffic control, and delivery~\cite{survey1}. Notwithstanding these several use case scenarios, one recent application attracted the researchers' attention worldwide. It consists of using drones as aerial base stations (ABSs) to provide coverage for ground users (GUEs). The latter scenario is believed to be compelling as it might provide interesting alternatives for users in an outage, for instance at gatherings or in the hardly accessible regions. During such scenarios, the classical cellular network architecture gets overwhelmed with multiple service requests, and hence UAVs intervene as rescuers. On this basis, literature related to deploying UAVs as ABS is gradually maturing with a wealth of publications in this area. On the one hand, several classical techniques have been used to propose overarching studies for the UAV deployment problem such as optimization and stochastic geometry~\cite{TrendsSG,IntegratingSG}. On the other hand, machine learning was also used to provide efficient solutions enabling a smart UAV deployment~\cite{mypaper2,MLsurvey}. Although deploying UAVs as ABSs seems to be tremendously exciting, its practical implementation faces myriad challenges that we cannot hide. Undoubtedly, the notable vulnerability to bad weather conditions is one of the downsides related to the use of drones as ABSs. Added to that the fact that drones must always keep a visual LOS with the controller, according to most of the latest drone aviation regulations. Furthermore, drones are known to be unable to process heavy tasks on board as their equipped hardware is usually limited. This could be explained by the limited payload that can be carried on commercial drones, which is irrevocably linked to the quality of hardware that can be installed onboard.  That being said, the UAV constrained energy remains undeniably the ultimate UAV drawback that will be our main focus in this work.\par
Broadly speaking, there have been several attempts in the literature to optimize the drone's endurance, especially for long-duration missions. To make it simple, we classify these solutions into (i) software-based solutions where mathematical solutions and algorithms have been proposed,  and (ii) hardware-based solutions where special devices have been designed to prolong the battery capacity. Starting with a software-based solution, we mention resource allocation and scheduling for UAV-based networks where authors in~\cite{RLUAV} proposed an energy-efficient resource allocation framework based on multi-agent reinforcement learning. In the same context, Sun et al.~\cite{SolarUAV} investigated resource allocation for solar-powered drones by maximizing the system sum throughput. Along with resource allocation, trajectory optimization also attracted researchers' attention. Authors in~\cite{Trajectory} proposed an energy-efficient UAV trajectory by maximizing the communication throughput and minimizing the UAV's energy consumption. Furthermore, battery optimization is also investigated in literature~\cite{Battery1,Battery2}. With this in mind, a great source of concern might arise when considering running such powerful solutions onboard UAV processors, especially with the present limited UAV resources. As an alternative to the previously mentioned solutions, a growing body of literature has examined equipping drones with physical solutions to lengthen their hovering time. Within this frame, we bring solar-powered drones to the reader's attention. As its name suggests, it consists of equipping drones with adequate solar panels to harvest daylight energy. Although several interesting research works targeted this area~\cite{SolarUAV,SolarPanel1,SolarPanel2}, this solution remains limited since it loses its efficiency at night. Added to that, the considerable complexity of equipping lightweight and energy-efficient solar panels onboard small drones. This is why other research publications targeted the use of tethered drones as an attempt to provide UAVs with indefinite battery life~\cite{tethered1,tethered2}. This obviously comes with the heavy cost of physically linking drones with the ground stations. That being said, laser light recently stole the limelight from tethered drones, as laser beams have been used to provide indefinite wireless charging for drones.\par
For many years, laser light was used primarily for applications related to surgery, industry, and especially communication~\cite{Trichili:20}. However, in recent decades, several companies have begun to use lasers to enable high-power beaming.  For instance, LaserMotive, also known as PowerLight Technologies, managed to win the NASA power beaming prize for their laser-powered robot. The same company succeeded in powering a drone for over 48 hours~\cite{lasermotive}. More recently, they held a demonstration of transferring more than 400W at a distance of 325m, which is indeed an exciting achievement. In addition to powering drones, high-power laser beaming has also been used to attack and eliminate drones. Companies such as Lockheed Martin and Boeing have developed several exciting projects in this area. The proposed system, usually referred to as laser beam director (LBD), consists of using a dedicated optical system formed by a set of mirrors to beamform the laser arrays and direct them towards their target. Although it is beyond the scope of this paper, within such a complex setup, it is reasonable to worry about safety measures. This is why several systems are put in place to guarantee safety such as the tracking systems, cooling systems, and eye-safety systems.\par
 In this paper, we analyze the performance of the deployment of laser-powered UAVs as ABSs to assist terrestrial base stations (TBSs) in providing coverage for GUEs. We use ground LBDs to provide energy to the UAVs, so that they can carry out long-duration missions. As proposed in our previous work~\cite{mypaper}, we assume simultaneous energy and information transfer to enable a backhaul link between the drone and its serving LBD.  That being said, we propose keeping a backup battery to be used in critical cases, when the drone has to leave its mission area or if by accident a LOS is lost.  To this end, we use tools from stochastic geometry to model the system and provide the expression of some performance metrics based on which we assess the system performance. In what follows, we review the most related works to the present paper and highlight the novelty and contribution of our work.
 \subsection{Related works}
 The use of stochastic geometry in system modeling is now a mature field where a surfeit of publications already exists. With the recent integration of ABSs into existing networks, researchers attempted to use tools from stochastic geometry to provide comprehensive studies and performance analysis for UAV-assisted networks. In this context, authors in~\cite{Low-altitude} used stochastic geometry tools to provide performance analysis for a network of UAVs modeled as a Poisson point process (PPP) for serving one user in an urban environment. Authors in~\cite{galkin2017stochastic} consider a network of UAVs and LTE base stations (BSs), both modeled with PPPs, where the BS provides the UAV with a backhaul link. Furthermore, in~\cite{Vertical2} a coverage and rate analysis has been proposed for a network of mixed TBSs and ABSs, modeled with PPPs and serving a typical user.  Moreover, Poisson cluster processes (PCPs) have been used to model the distribution of users on the ground. In \cite{added1},  authors used Thomas cluster process and Mat\'ern cluster process (MCP) to model the distribution of users, while in \cite{Abovehotspot}, users were modeled with  MCP where the drones are positioned above the cluster centers. PPP is known to be tractable and easy to exploit compared to other point processes, however, it assumes that the user's position is completely random. Compared to PPP, MCP is considered more realistic in scenarios where the distribution of users is described by gathering groups, called clusters. This being said, MCP requires knowledge of the statistics of these clusters to adjust the parameters of the point process such as the cluster radius.  Furthermore, authors in~\cite{downlink} used binomial point processes (BPP) to model the distribution of the drones. In the same context, authors in~\cite{Vertical} model the drones with a BPP, consider the existence of  TBSs, and provide a downlink coverage and rate analysis. None of the previously mentioned research works have focused on the energy aspect of drones which is a crucial factor, especially when deploying UAVs as ABSs.\par
 Compared to the use of stochastic geometry in modeling, the use of laser beams for drones is still under investigation. Authors in~\cite{throughput} studied the downlink throughput maximization problem for a laser-powered UAV.  The considered fixed-wing drone in this work uses the laser beam to harvest enough energy for hovering and for communication with the ground terminal. Moreover, authors in~\cite{IoT} consider the use of laser-powered drones in assisting an Internet-of-Things (IoT) network. An optimization problem is formulated to maximize the number of IoT devices connected with the drone. In~\cite{laser3} the problem of joint power and trajectory optimization was studied for rotary-wing laser-powered drones.   In \cite{r1} authors studied a UAV-aided multicasting system for simultaneous FSO backhaul and power transfer. Under PPP distributed users,  the UAV altitude and the FSO and RF transmissions are optimized. Statistics of the FSO communication channel between a UAV and a central unit were proposed in \cite{r2}. Channel models were derived taking into account the position and orientation fluctuation of the drone as well as the non-orthogonality of the laser beam.  In another context, Jaafer et al. provided a battery dynamics study for a laser-powered drone under different charging techniques~\cite{Battery2}. Authors in~\cite{adressingbattery} investigated different charging techniques for drone networks including laser-powered drones. These research works do not provide any performance analysis for BS-mounted drones. We provided a stochastic geometry-based analysis in our previous work~\cite{mypaper}. We studied the communication coverage and energy coverage of a single UAV powered by a network of ground LBDs, however, we did not consider using multiple drones, multiple users, or multiple TBSs. 
 \subsection{Novelty and Contributions}  
All of the above-mentioned works either consider applying stochastic geometry for UAV-assisted networks or studying laser-powered drones from other angles rather than stochastic geometry, whereas in our work:
\begin{itemize}
	\item We design a complex system setup, composed of a network of TBSs, multiple UAVs, multiple GUEs, and a network of LBDs. We use various point processes to model their spatial distribution. 
	\item  We propose and analyze a novel deployment strategy for the laser-powered drones that enables a trade-off between energy and wireless coverage. 
	\item We derive the theoretical expressions of the coverage probability, the Laplace transform of interference, and the coverage probability. We validate these expressions through numerical simulations in different types of environments.
	\item Finally, we provide high-level insights about the proposed system based on the derived performance metrics.
\end{itemize}   
{\em Notations:} In this paper, we use $\mathbb{E}(\cdot)$ to denote the expectation operator and $\mathbb{P}(\cdot)$ for the probability of an event. $\mathds{1}(\cdot)$ denotes the indicator function while  $\delta(\cdot)$ is the Dirac delta function, equal to one at zero and null otherwise. Function $F_x(\cdot)$ denotes the cumulative distribution function (CDF) of a given random variable $x$. Set $\mathbb{R}_+$ denotes  non-negative real numbers and $\mathbb{R}^2$ denotes the set of 2-dimensional vectors with real-entries. For $x \in \mathbb{R}^2$, $||x||$ denotes the euclidean norm of $x$. We denote by $|\cdot|$ the absolute value operator.
\vspace{-3mm}
\section{System Model}\label{Sec:SystemModel}
\begin{figure}
	\centering
	\captionsetup{justification=centering}
	\includegraphics[width=3.5in]{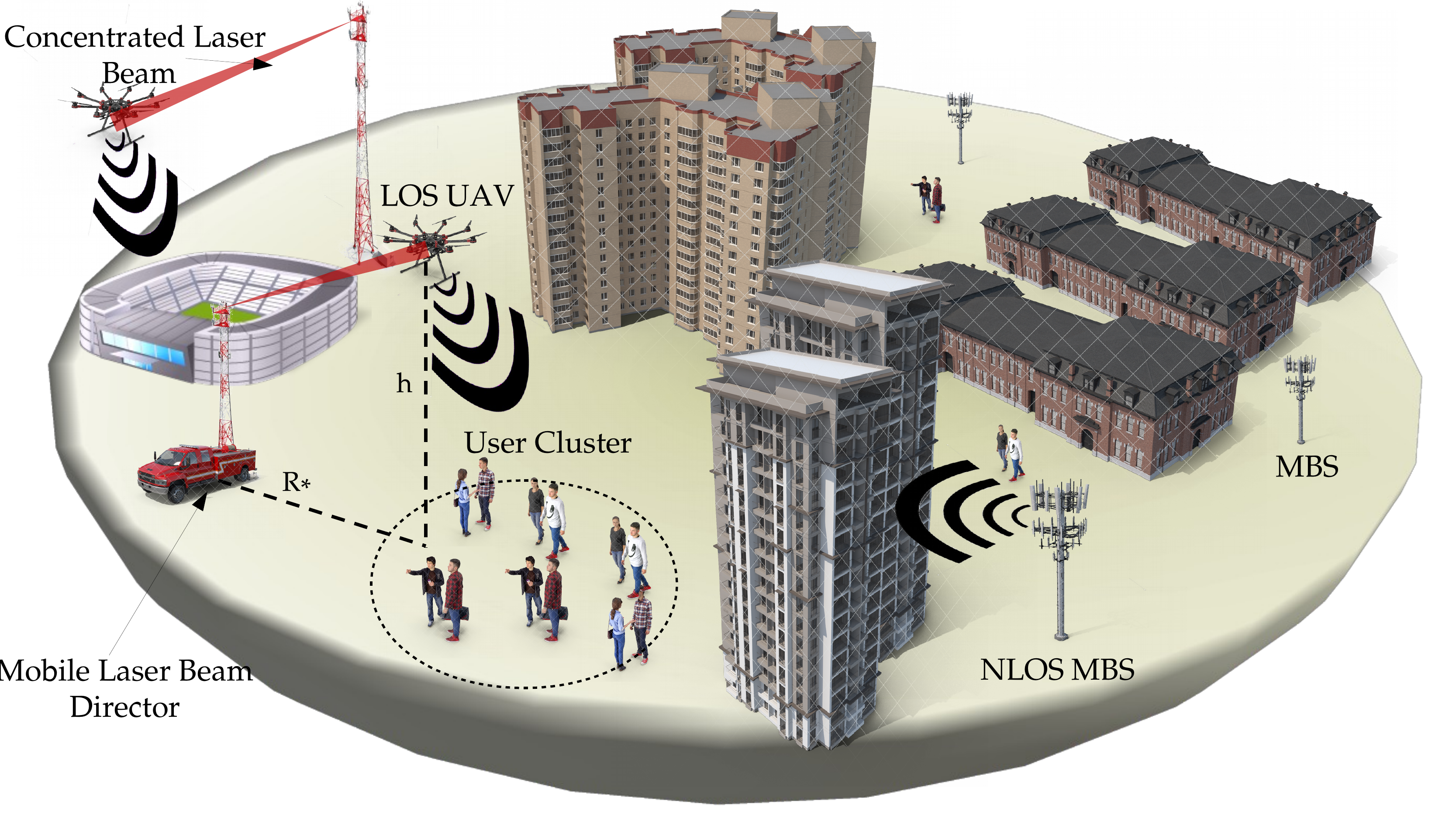}
	\caption{Laser-powered UAV network.}
	\label{fig:UseCase}
\end{figure}
In this work, we consider the scenario of assisting a poor terrestrial TBSs architecture in providing wireless coverage to GUEs through the use of drones that are acting as ABSs. The UAVs, while performing their task, benefit from the energy supply provided by the terrestrial LBDs that are statically placed on the ground. Figure~\ref{fig:UseCase} represents one use case for laser-powered drones used to provide coverage for GUEs. Such users may be located in small geographic areas, which motivates modeling them with clustered processes. Figure~\ref{fig:UseCase} also hints at the fact that the deployment of LBDs can be either static or dynamic. Moreover, the LBDs guarantee energy coverage for the drones as long as the UAVs are placed within an adequate range $R^*$ that is derived later. Consequently, the drones will not rely on their battery but use them only in critical situations, allowing them to carry out long-duration missions. We assume that an LBD can only serve one UAV at a time, so temporal coordination is beyond the scope of this paper. Therefore, the case where more than one UAV requires recharging is included in the critical situation we have mentioned previously.  The environments that we target in this study are suburban, urban, and dense urban, where the distributions of buildings may affect the performance analysis significantly. In what follows, we present the different models used in this work.

\subsection{Spatial Distributions}
We assume that we do not have any prior knowledge about the optimal positions of the LBDs and model their distribution with a homogeneous PPP, denoted by $\Phi_{LBD}$, with corresponding density $\lambda_{LBD}$. Consequently, our performance results can serve as a lower bound since we do not optimize the LBD locations. We also consider the existence of a network of TBSs that is modeled with another homogeneous PPP, denoted by $\Phi_{b}$, with density $\lambda_{b}$. Taking advantage of the nature of the users' gathering, that usually forms random groups creating a congesting in the wireless network traffic, we model the GUEs distribution with a MCP where the distribution of the parent nodes, referring to the cluster center, forms a PPP denoted by $\Phi_{cc}$ with density $\lambda_{cc}$ and maximum cluster radius $r_{ max}$. For the moment, we denote the UAV distribution by $\Phi_{u}$ with density $\lambda_{u}$,  where $x_i$ is the position of the $i^{th}$ UAV. We explain, later in this work, our proposed deployment strategy for the drones.
\begin{figure}
	\centering
	\captionsetup{justification=centering}
	\includegraphics[width=4.5in]{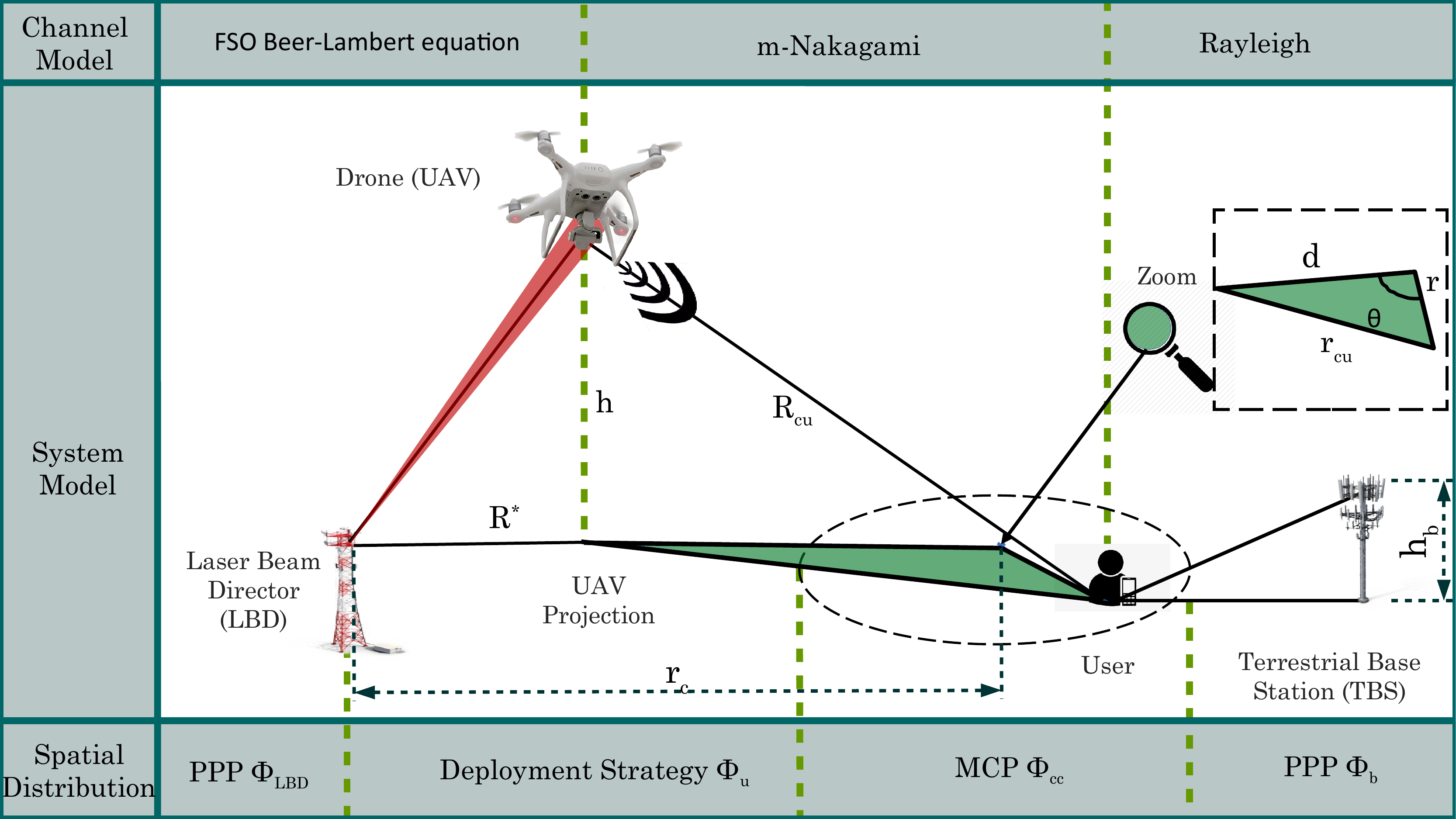}
	\caption{System model summary.}
	\label{fig:SystemModelSummary}
\end{figure}
\subsection{LOS Blockage Probability Model}
It is crucial in our work to identify the link type between the transmitter and the receiver. To this end, we model the probability of LOS link such that it captures the effect of the distribution of the surrounding buildings within an urban environment. Various complex models were proposed in the literature~\cite{LOSProbabilityComplicated}, however, in our work, we use the simplified LOS probability, derived in~\cite{Vertical}, to model the link UAV-user and TBS-user:                     
\begin{gather}
	\mathcal{P}_{\rm i}(r)= -a\; \exp\left(-b\; \arctan\left(\frac{h_i}{r}\right)\right)+c, \forall i \in \{Lu,Lb \},\\
	\mathcal{P}_{\rm Nu}(r)=1-\mathcal{P}_{\rm Lu}(r),\\
	\mathcal{P}_{\rm Nb}(r)=1-\mathcal{P}_{\rm Lb}(r),
\end{gather}
where $\mathcal{P}_{\rm Lu}$, $\mathcal{P}_{\rm Nu}$, $\mathcal{P}_{\rm Lb}$ and $\mathcal{P}_{\rm Nb}$ represent respectively the UAV LOS, UAV non-LOS (NLOS), TBS LOS, and TBS NLOS probabilities,  $r$ is the horizontal distance to the respective node,  $h_{\rm Lu}= h_{\rm Nu}=h$ is the UAV altitude, $h_{\rm Lb}= h_{\rm Nb}=h_b$ is the TBS height.  The parameters $a$, $b$, and $c$ depend on the environment nature. 
\subsection{PPP Thinning Procedure}
In this section, we derive the different distances to the nearest LOS UAV, NLOS UAV, LOS TBS, and NLOS TBS. To this end, a common technique in dealing with LOS and NLOS distributions of a given point process is to take advantage of the thinning operation in dividing the parent PPP into two non-homogeneous PPPs. The operation consists of deleting each point with a given probability. In our case, we use $\mathcal{P}_{\rm Lu}(r)$ in dividing $\Phi_u$ into a set of LOS UAVs denoted by $\Phi_{\rm Lu }$ and a set of NLOS UAVs, denoted $\Phi_{\rm Nu }$. The same is done to $\Phi_b$ by dividing it into $\Phi_{\rm Lb }$ and $\Phi_{\rm Nb }$. The density of each point process is given by: 
\begin{align}
	\lambda_{i}(r)= \lambda_{\rm u} \mathcal{P}_{\rm i}(r), i \in \{\rm Lu,Nu\},
	\\	\lambda_{i}(r)= \lambda_{\rm b} \mathcal{P}_{\rm i}(r), i \in \{\rm Lb,Nb\}.
\end{align} 
\subsection{Channel Modeling}
The UAV-to-user channel is modeled by a Nakagami-$m$ multi-path fading denoted by $G_{\rm Lu}$ and $G_{\rm Nu}$. However, the terrestrial BS-to-user channel is modeled by a Rayleigh fading denoted by $G_{\rm Lb}$ and $G_{\rm Nb}$. We use this notation to provide a compact notation for the equations later. The instantaneous power received from a node placed at $x_{\rm i}$ at the reference user is given by: 

\begin{align}
\mathcal{R}_{\rm j}(r(x_i))=\rho_{\rm j}\; \eta_{\rm j} \;\left(r(x_i)^2+h_j^2\right) ^{-\frac{\alpha_{\rm j}}{2}} G_{\rm j} \textit{ , } x_i \in \Phi_{\rm j},  \textit{ and } j \in \mathcal{C},
\end{align}
where $\rho_{\rm Lu}=\rho_{\rm Nu}=\rho_{\rm u}$  denotes the UAV transmit power, $\rho_{\rm Lb}=\rho_{\rm Nb}=\rho_b$  denotes the TBS transmit power, $\eta_{\rm i}$ represents the additional path loss related to a LOS condition or NLOS conditions, and $r(x_{\rm i})$ is the horizontal distance to the point located at $x_{\rm i} \in \mathbb{R}^2$. Hereinafter, we denote by $\mathcal{C}$ the set $\{ \rm Lu,Nu,Lb,Nb\}$. For the ease of notation, we also define the average power received from a node place at $x_i$ at the reference user as follows:
\begin{align}
\overline{\mathcal{R}}_{\rm j}(r(x_i))=\rho_{\rm j}\; \eta_{\rm j} \;\left(r(x_i)^2+h_j^2\right) ^{-\frac{\alpha_{\rm j}}{2}} \textit{ , } x_i \in \Phi_{\rm j},  \textit{ and } j \in \mathcal{C}.
\end{align}

\subsection{Energy Harvesting and Backhaul Link}
As proposed in our previous work~\cite{mypaper}, we consider simultaneous communication and energy transfer under a power splitting technique at the drone level, where the LBD at the ground does not only guarantee energy supply but also provides a backhaul link to the drones.  
To model the energy transfer link between the drones and the LBDs, we use the following model based on the Beer-Lambert equation~\cite{killinger}: 
\begin{align} \label{eq:EnergyHarvesting}
	p_{\rm harv}(R)=h_t \frac {\left(1-\delta_s\right) \omega T\chi \;  p_{\rm trans} \exp\left(-\alpha \sqrt{R^2+h^2}\right)}{\left(D+\sqrt{R^2+h^2}\Delta\theta\right)^{2}},
\end{align} 
where $R$ is the horizontal distance to the drone,  $h_t$ is a log-normal random variable representing the turbulence effect,  $\delta_s$ is the power splitting factor used to separate the energy transfer link and the communication link. The parameter $w$ denotes the receiver conversion efficiency, $T$ is the area of the receiver telescope or collection lens, $D$ is the size of the initial laser beam, $\alpha$ is the attenuation coefficient of the medium, $\chi$ is the combined transmission receiver optical efficiency, and $\Delta\theta$ is the angular spread of the laser beam. The angular spread $\Delta\theta$ is equal to $\frac{D_d}{f}$, where $D_d$ is the size of the detector and $f$ its focal length.\par 
Regarding the backhaul link, we propose using a basic laser intensity modulation such as on-off keying for communication. Accordingly, the signal-to-noise ratio (SNR) level at the UAV is given by \cite{agrawal2012fiber,killinger}: 
\begin{align}
{\rm SNR_{\rm UAV}}(R) =\frac{\delta_s\; \eta\; h_t\; \omega A\chi \; p_{\rm trans} \;\exp\left(-\alpha \sqrt{R^2+h^2}\right)}{2 \;\hbar\; \nu\; \Delta f \left(D+\sqrt{R^2+h^2}\Delta\theta \right)^{2}},
\end{align}
where $\hbar$ is Planck's constant, $\nu=\frac{c}{\lambda}$ is the photon's frequency, $\lambda$ is the wavelength, $c$ is the speed of light, and $\Delta f $ is the modulation frequency bandwidth.\par
\section{UAV Deployment strategy}
In this section, we present the details of our proposed UAV deployment. We consider it a large-scale UAV deployment strategy since we use a PPP to model multiple drones and simulate them in a large area of interest. In what follows, we derive the critical charging distance $R^*$, then we use its value to define our deployment strategy. Finally, we identify the resulting UAV distribution after applying the proposed placement strategy.
\subsection{Critical Charging Distance} \label{subsec:CriticalChargingDistance}
To guarantee a safe energy level and successful communication with the LBD, the UAV flies inside a ball centered around the serving LBD and of radius $R^*$. This critical charging distance was the subject of~\cite{mypaper}, where we provided its  closed-form expression in the non-turbulent \footnote{Refers to the case where the effect of the optical turbulence on energy harvesting is not taken into consideration.  } regime as follows: 
\begin{align}
	\medmath{R^*=\Bigg[ \Bigg\{\frac{2}{\alpha}W_0\left(\frac{\alpha}{2\Delta\theta }\sqrt{\frac{\left(1-\delta_s\right)\omega T\chi  p_{\rm trans} e^{\frac{\alpha D}{\Delta\theta }}}{p_{\rm prop}+p_{\rm comm}}}\right)-\frac{D}{\Delta\theta}\Bigg\}^2-H^2\Bigg]^{\frac{1}{2}},}
\end{align}
where $W_0(.)$ is the principal branch of the Lambert W function~\cite{corless1996lambertw}, $P_{\rm prop}$ is the UAV propulsion power, and $P_{\rm comm}$ is the UAV communication power. However, the laser intensity is usually affected by optical turbulence, especially for long range applications.

\begin{lemma}[Critical charging distance] \label{lemma: CriticalChargingDistance} In the Case of log-normal turbulence effect, where  $h_t\sim \lognormal(-2\sigma,2\sqrt{\sigma})$, the value of the critical charging distance $R^*$, is derived numerically by solving the below equation:
	\begin{equation}
		\label{eq:Rstar}
P_{\rm harv}(R^*)\exp \left(2\sqrt{2\sigma(R^*)} \erf^{-1}(1-2\mathcal{B}) -2\sigma(R^*)\right)-p_{\rm prop}-p_{\rm comm}=0,
	\end{equation}
where $R^*$ is  derived such that the joint SNR and energy coverage  probability is above a certain level denoted by $\mathcal{B}$.
\end{lemma}
\begin{IEEEproof}
	See Appendix~\ref{app:1}.
\end{IEEEproof}
	We bring to the reader's attention the following points: (i) $\sigma$ is denoted by $\sigma(R^*)$ in (\ref{eq:Rstar}) to emphasize the fact that it is a function of $R^*$ and that a closed-form solution could not be easily derived since a commonly used expression is $\sigma(R^*)=\sqrt{0.3 k^{\frac{7}{6}} C^2_n(h){R^*}^{\frac{11}{6}} }$, where $C^2_n(h)$ is the refraction index at an altitude h, and k is the wavenumber. Moreover, $p_{\rm harv}(R^*)$ is not linear in $R^*$, as  (\ref{eq:EnergyHarvesting}) shows, which complicates further a closed-form solution. (ii) Typical values for the minimal threshold $\mathcal{B}$ are [0.9, \;0.99], which means that successful communication energy coverage is guaranteed with probability 90\% to 99\%. Regarding the outage probability, we rely on the drone's backup battery to get back to the LBD's field of view or to a safe-charging range. 
\begin{table}[]\caption{System notations}
	\label{Table1}
	\centering
	\resizebox{\columnwidth}{!}{%
	\begin{tabular}{|c|c|c|c|}
		\hline
		\multicolumn{4}{|c|}{\textbf{Notations}}                                                                                                              \\ \hline
		\textbf{Notation} & \textbf{Description}                                    & \textbf{Notation}   & \textbf{Description}                              \\ \hline
		\multicolumn{2}{|c|}{\textbf{Distribution notations}}                       & $\delta_s$          & Power splitting factor                            \\ \hline
		LBD               & Laser beam director                                     & $\alpha$            & Attenuation coefficient of the medium             \\ \hline
		Lu                & LOS UAV                                                 & $\Delta \theta$     & Angular spread of the beam                        \\ \hline
		Nu                & NLOS UAV                                                & $w$                 & Receiver conversion effiency                      \\ \hline
		Lb                & LOS TBS                                                 & $T$                 & Area of the receiver telescope or collection lens \\ \hline
		Nb                & NLOS TBS                                                & $k$                 & Wavenumber                                      \\
 \hline
		\multicolumn{2}{|c|}{\textbf{Distribution-dependent notations}}             &  $\chi$              & Combined receiver optical efficiency                                       \\ \hline
		$\Phi_{i} $       & Distribution $i$                                        &      $p_{harv}$          & Harvested power at the UAV            \\ \hline
		$\rho_i$             & Transmit power                                          &     $p_{\rm prop}$       &     Propulsion power for the UAV                     \\ \hline
		$\eta_i$          & Additional path loss                                    & $p_{\rm comm}$      &  Communication power for the UAV                      \\ \hline
		$\alpha_i$        & Path loss                                               &  $\sigma(R^*)$      &    Turbulence standard deviation              \\ \hline
		$\lambda_i$       & Density                                                 &   $D$            &     Size of the initial laser beam                                               \\ \hline
		$r_i$             & Distance to the nearest neighbor from $\Phi_i$          & \textbf{$C^2_n(h)$}          &   Refraction index at altitude h                     \\ \hline
		$G_i$             & Nakagami-m random multipath fading                  & \multicolumn{2}{c|}{\textbf{Other notations}}                                        \\ \hline
		$m_i$             & Nakagami-m fading term                     & $h$ &   UAV altitude                      \\ \hline
		$h_i$             & Altitude of distribution $\Phi_i$ nodes &     $h_b$             &      TBS altitude                              \\ \hline
		$\mathcal{P}_i$   & Probability of LOS                                      &  $r_{\rm max}$             &    Cluster radius                                    \\ \hline
		\multicolumn{2}{|c|}{\textbf{Laser link notations}}                         & $\hbar$       &  Planck's constant                                 \\ \hline
		$R^*$             & Critical charging distance                              &$\mathcal{S}$           &      Deployment strategy                                            \\ \hline
		$\mathcal{B}$               & Joint energy-communication coverage threshold           & $\mathcal{C}$          &  The set $\{ \rm Lu, Nu,Lb,Nb\}$                             \\ \hline
	\end{tabular}
}
\end{table}
\subsection{UAV Deployment Strategy Description}\label{subsec:DeploymentStrategyDescription}
\begin{figure}
	\centering
	\captionsetup{justification=centering}
	\includegraphics[width=5in]{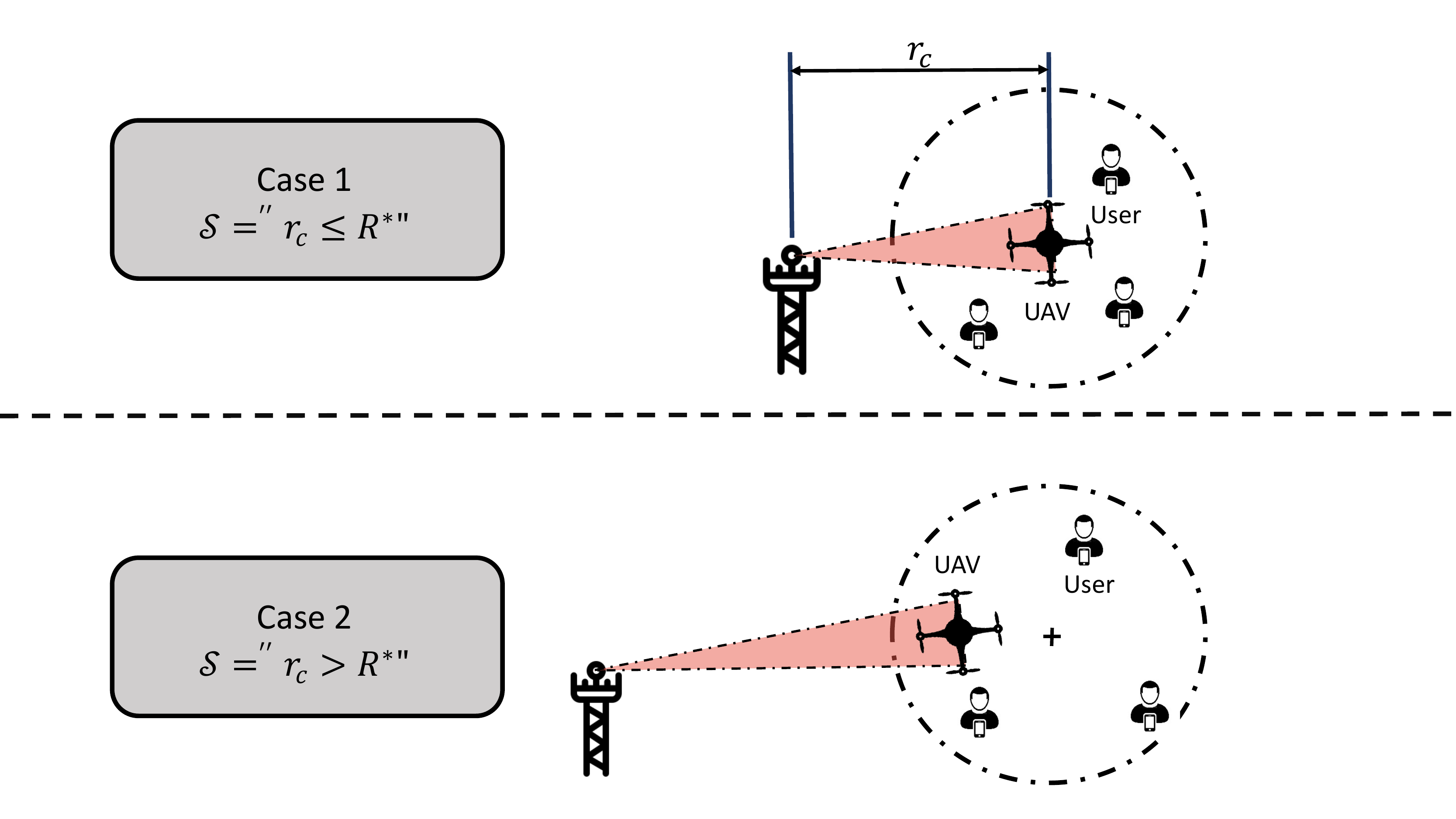}
	\caption{Deployment strategy overview.}
	\label{fig:DeploymentStrategy}
\end{figure}
Positioning the UAV above the cluster centers might provide good coverage for the users, but can not be optimal in terms of energy coverage as it is not guaranteed that the distance to the serving LBD is always in safe ranges. Based on the distance $R^*$, we propose a deployment strategy, denoted by $\mathcal{S}$, described by Figure~\ref{fig:DeploymentStrategy} and Algorithm~\ref{algorithm},  that establishes a trade-off between energy coverage and the quality of service provided. Below, we summarize the main points of the UAV placement strategy:
\begin{itemize}
		\item One UAV is dedicated to each user cluster center. 
		\item Case1.	$\mathcal{S}=`` r_c \leq R^*"$: If the distance from the cluster center to its nearest LBD is below $R^*$, the drone decides to hover above the cluster center to provide maximum coverage for the users inside the cluster center. This is assumed to provide the lowest average euclidean distance to the users inside the cluster and thus results in better coverage. 
		\item Case2.	$\mathcal{S}=`` r_c  > R^*"$: If the distance from the cluster center to its nearest LBD is above $R^*$, The UAV is forced to shift along the line linking the cluster center and its nearest LBD. This is because the harvested energy level at the cluster center cannot compensate for the consumed energy. Thus the UAV is placed at a distance $R^*$ on the line linking the drone from its nearest LBD.
		\end{itemize}
\begin{algorithm}
	\caption{UAVs Relocation  Algorithm}\label{algorithm}
	\begin{algorithmic}[1]
		\Procedure{Relocate}{$\Phi_{LBD},\Phi_{cc}$}
		\State $R^* \gets \textit{solve(\ref{eq:Rstar})} $
		\State $\Phi_{u} \gets \Phi_{cc}$
		\For{$(x,y) \in \Phi_{u}$} 
		\For{$(x',y') \in \Phi_{LBD}$} 
		\State $d(x',y') \gets \sqrt{(x-x')^2 +(y-y')^2}$
		\EndFor
		\State $r_c \gets $min($d$)
		\State $(x_n,y_n) \gets $argmin($d$) \Comment{$(x_n,y_n)$ is the nearest neighbor of (x,y) from $\Phi_{LBD}$}
		\State slope $\gets \frac{y-y_n}{x-x_n}$
		\If{$(r_c\geq R^*)$ }\Comment{set (x,y) such that the distance to the nearest LBD is set to $R^*$} \State $x\gets x_n+\frac{{\rm sign}(x-x_n)R^*}{\sqrt{1+slope^2}}$
		\State $y\gets y_n+\frac{{\rm sign}(y-y_n)R^*}{\sqrt{1+slope^2}}$\Comment{In case $r_c<R^* $ keep (x,y) unchanged.}
		\EndIf
		\EndFor
		\State \textbf{return} $\Phi_{u}$
		\EndProcedure
	\end{algorithmic}
	
\end{algorithm}
\vspace{-4mm}

\subsection{UAV distribution}
\begin{figure}
		\centering
		\captionsetup{justification=centering}
		\includegraphics[width=4in]{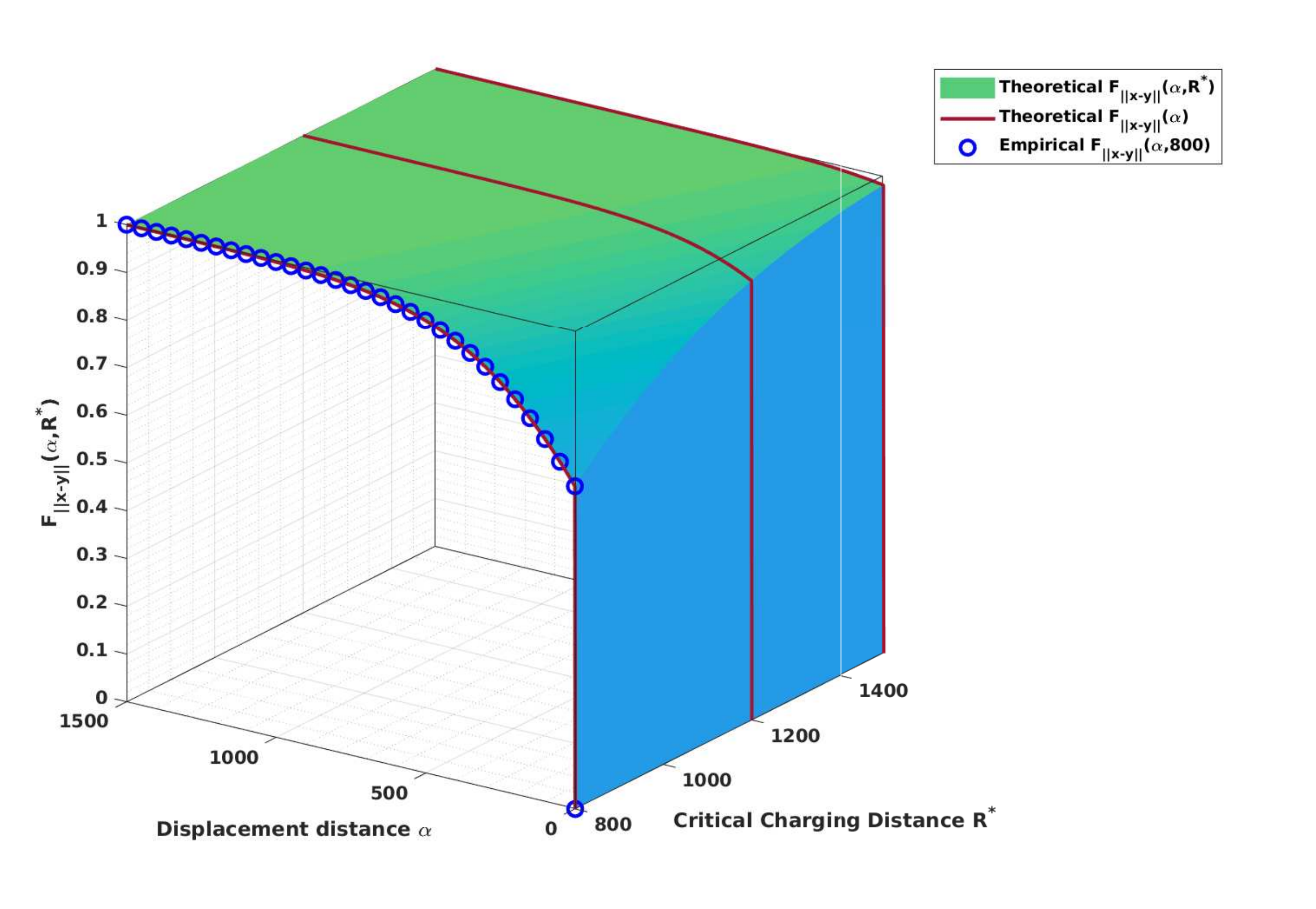}
		\caption{Displacement magnitude as a function of $R^*$ and the displacement distance $\alpha$, $\lambda_{LBD}=0.52\times10^{-6}$ unit/m$^2$ and $\lambda_{u}=3\times10^{-6 }$ unit/m$^2$.}
		\label{fig:DisplacementNorm}
	\end{figure}

In this subsection, we present the resulting UAV distribution after applying Algorithm \ref{algorithm}. Applying random translations to a PPP only affects its density, and the resulting density is provided in~\cite{haenggi}. However, in our case, we cannot apply the displacement theorem to derive the new density. This is because the translations we are making are not independent. In this case, and from the reference user point of view, we consider the existence of a central UAV, which is the drone associated with the cluster of the reference user and we approximate the rest of the UAV distribution as a PPP. We show later through the numerical simulation that this approximation is decent. In what follows, we provide the ranges of the critical charging distance at which this approximation is decent. To do so, we analyze the  displacement magnitude, which is the distance crossed by a drone following the placement strategy. Let $x,y \in \mathbb{R}^2$, where $x \in \Phi_u$ denotes the location of one UAV and y denotes its new position after the relocation described in Algorithm \ref{algorithm}. Let $\mathcal{NN}(x)  \in \Phi_{LBD}$ be the nearest LBD to $x$. Notice that $|| x-y||$ is a mixed random variable from the following PDF:
\begin{align}
	f_{||x-y||}(\alpha)= \left(1-\exp\left(-\pi \lambda_{\rm LBD} {R^*}^2 \right)\right)\delta(\alpha) +2 \lambda_{\rm LBD} \pi (R^*+\alpha) \exp\left(-\pi \lambda_{\rm LBD} (R^*+\alpha)^2\right) \mathds{1}(\alpha \geq 0).
\end{align}
This is because the CDF of $||x-y||$ is derived as follows: 
\begin{align}
	F_{||x-y||}(\alpha)&= \mathbb{P}\left(||x-y|| \leq \alpha\right)\nonumber, \\
	&=\mathbb{P}\left( || \mathcal{NN}(x) -x || \leq R^*+\alpha \right)\nonumber, \\
	&=  1-\exp\left(-\pi \lambda_{\rm LBD} {(R^*+\alpha)}^2 \right).
\end{align}
Thus, the displacement magnitude $|| x-y||$ is a mixed random variable with a continuous part: 
\begin{align}
	C(\alpha)=
	\begin{dcases} 
		\exp\left(-\pi \lambda_{\rm LBD} {R^*}^2 \right)-\exp\left(-\pi \lambda_{\rm LBD} {(R^*+\alpha)}^2 \right)& \alpha \geq 0 \\
		0 & else.
	\end{dcases}
\end{align}
Accordingly, the discrete part which discribes the probability of not displacing the node $x$, is given by: 
\begin{align}
	D(\alpha)=
	\begin{dcases} 
		1-\exp\left(-\pi \lambda_{\rm LBD} {R^*}^2 \right)& \alpha \geq 0 \\
		0 & else. 
	\end{dcases}
\end{align}
In Figure \ref{fig:DisplacementNorm}, we plot the theoretical CDF of the displacement norm versus the empirical CDF for different ranges of the critical charging distance $R^*$ and the displacement distance $\alpha$. Based on this figure, we can notice that for reasonable ranges of the critical charging distance $[900m\; \;1300m]$, the probability of displacement is low and the nodes are often not relocated. This is represented by the blue-colored facet of the cube. Moreover, even if the node is relocated then the displacement distance is not significant as shown by the green-colored facets of the cube in Fig~.\ref{fig:DisplacementNorm}. 

\section{Performance Analysis}

To assess the performance of the system, we start by deriving the important distance distributions, the necessary association probabilities, the Laplace transform of interference, and finally the coverage probability.
\subsection{Important Distance Distributions} 

\subsubsection{Distance to the Nearest Neighbor from a PPP}
The distance distribution to the nearest point from any PPP is given in~\cite{haenggi} by: 
\begin{align}
	f_{r_{\rm i}}(r) = 2 \pi\; \lambda_{i}(r)\; r \;\exp\left(-2 \pi \int_{0}^{r} \lambda_{i}(x) x \;dx\right), 
\end{align}
where $r_i$ is the distance to the nearest neighbor from $\Phi_i$, with corresponding density $\lambda_i$, and such that  $i \in \{Lu,Nu,Lb,Nb,LBD\}$.
\subsubsection{Distance Distribution Between the Reference User and its Central UAV}
Following the deployment strategy that we proposed previously in Sec.~\ref{subsec:DeploymentStrategyDescription}, it is clear that a single UAV is associated with each user cluster. From the reference user point of view, the UAV associated with its cluster center is considered as the central UAV. Deriving the distance distribution between the reference user and its central UAV, denoted by $r_{\rm cu}$, is important as it is used later in deriving the association probabilities and finally the coverage probability. As indicated in Algorithm~\ref{algorithm}, this distance depends on $r_c$, which is the distance from the cluster center to the nearest LBD.\\
\textbf{Case1. Central UAV is placed above the cluster center ($r_c \leq R^*$)}\\
In this scenario, as the UAV is placed above the cluster center to maximize its coverage for GUEs, and since the users are uniformly distributed inside the cluster of radius $r_{max}$, the distance distribution to the reference user is given by:  
\begin{gather}
 f_{\rm r_{cu}|r_c \leq R^*}(r)= \frac{2r}{r_{max}^2} \mathds{1}\left(0 \leq r \leq r_{max} \right),
\end{gather}
where $r_{max}$ is the maximal cluster radius.

\textbf{Case2. Central UAV is placed at a distance $R^*$ from its nearest LBD ($r_c > R^*$)}\\\label{Sec:frcu}
In the case where being placed above the hotspot center does not guarantee enough energy level for the drone, the UAV is obliged to be placed at the distance $R^*$ from its nearest LBD.
\begin{prop}
	\label{Proposition:PDFrcu}
	For $ \alpha \in \mathbb{R}_+$, the PDF of the distance between the UAV and a reference user in the case where $r_c > R^*$, denoted by $f_{r_{\rm cu|r_c > R^* }}(\alpha)$, is given by:
	\begin{gather}
	\resizebox{0.9\textwidth}{!}{$	\begin{dcases} 
			f_{\rm r_{\rm cu}|r_c \leq R^* }\left(\alpha-r_{cc}\right) & |\alpha-r_{\rm cc}| \geq r_{max} \\ \\
			f_{r_{\rm cu}|r_c \leq R^* }\left(\alpha-r_{\rm cc}\right)-\frac{2\left(\alpha-r_{\rm cc}\right) \arcos\left(f\left(|\alpha-r_{\rm cc}|,\alpha\right)\right) }{\pi r_{max}^2} & |\alpha-r_{\rm cc}|<r_{max}\leq r_{\rm cc}+\alpha  \\+ \int_{|\alpha -r_{\rm cc}|}^{ r_{max}}\frac{2 \alpha}{ \pi r_{max}^2 r_{\rm cc} \sqrt{1-f^2(r,\alpha)}}dr & \textit{ and } \alpha \neq r_{\rm cc}  \\ \\
			\int_{0}^{ r_{max}}\frac{2 \alpha}{ \pi r_{max}^2 r_{\rm cc} \sqrt{1-f^2(r,\alpha)}}dr &   |\alpha-r_{\rm cc}|<r_{max}\leq r_{cc}+\alpha  \\ & \textit{ and } \alpha = r_{\rm cc} \\
			f_{r_{\rm cu}|r_c \leq R^* }(\alpha-r_{\rm cc})-\frac{2(\alpha-r_{\rm cc}) \arcos(f(|\alpha-r_{\rm cc}|,\alpha)) }{\pi r_{max}}& |\alpha-r_{\rm cc}|<r_{\rm cc}+\alpha <  r_{max}\\+ \int_{|\alpha -r_{\rm cc}|}^{ \alpha +r_{\rm cc}}\frac{2 \alpha}{ \pi r_{max}^2 r_{\rm cc} \sqrt{1-f^2(r,\alpha)}}dr+\frac{2(\alpha+r_{\rm cc}) \arcos(f(\alpha+r_{\rm cc},\alpha)) }{\pi r_{max}^2}  & \textit{ and } \alpha \neq r_{\rm cc}  \\ \\
			\int_{0}^{ \alpha +r_{\rm cc}}\frac{2 \alpha}{ \pi r_{max}^2 r_{\rm cc} \sqrt{1-f^2(r,\alpha)}}dr
			+\frac{2(\alpha+r_{\rm cc}) \arcos(f(\alpha+r_{\rm cc},\alpha)) }{\pi r_{max}^2} & \textit{else,}
		\end{dcases}$}
	\end{gather}
	
	where $r_{cc}=r_c-R^*$, and $f(r,\alpha) = \frac{(r_c-R^*)^2+r^2-\alpha^2}{2 (r_c-R^*) r}.$
\end{prop}
\begin{IEEEproof}
	See Appendix~\ref{app:2}.
\end{IEEEproof}

\begin{remark}
	To speed up the integration in the simulation, it is important to identify the support of the PDF. The support of the PDF $f_{r_{\rm cu} }(\alpha)$ is itself random and depends on the placement strategy $\mathcal{S}$ and consequently on the random variable $r_c$. However, when conditioning on the placement strategy, the support of the PDF of $r_{\rm cu}$ is given by  $[(r_c-R^*-r_{\rm max})^+,r_{\rm max}+ (r_c-R^*)^+ ]$, where $x^+$ denotes $\max(0,x)$. This can be justified as follows:
	\begin{itemize}
		\item 	The first case corresponds to placing the UAV above the cluster center ($r_c \leq R^*)$. Based on the properties of MCP the support of the PDF is $[0,r_{\rm max}]$.
		\item 	The second case corresponds to relocating the UAV according to Algorithm ~\ref{algorithm} and placing it inside the reference cluster as shown in Figure~\ref{fig:DeploymentStrategy}, which is mathematically explained by ($0<r_c-R^* \leq  r_{\rm max})$. In this case, we can show that the support of the PDF function is $[0,r_{\rm max}+r_c-R^*]$.
		\item The third case corresponds to relocating the UAVs but placing it outside the reference cluster  i.e.  $(r_{\rm max} < r_c - R^*)$. In this case we can show that the support of the PDF is $[r_c-R^*-r_{\rm max}, r_{\rm max}+r_c-R^*]$.
	\end{itemize}	
\end{remark}
\subsubsection{Distance to the Nearest Interferes}

The reference user associates towards the nearest point from each distribution since it provides, by definition, the strongest signal.
Since we use multiple distributions and in order to avoid any confusion, we provide  a compact form for the distance to the nearest interferer from each distribution. 

\begin{lemma}[Distance to the nearest interferer]
	\label{lemma:DistanceInterferer}
	Suppose that the reference user is connected to the nearest point from a distribution $\Phi_j$, distant by $r \geq 0$. The distance to the nearest interferer from the distribution $\Phi_i$ is  given by:
	\begin{align}
	f_j^{i}(r)= \sqrt{\max\left( 0, \left(\frac{\rho_i \eta_i}{\rho_j \eta_j}\right)^{\frac{2}{\alpha_i}}\left(h_j^2+r^2\right)^\frac{\alpha_j}{\alpha_i} -h_i^2\right)} \textit{ , } i,j \in \mathcal{C},
	\end{align} 
	where $\rho_i$ is the transmit power, $\eta_i$ is the additional path loss, $\alpha_i$ is the path loss exponent, and $h_i$ is the height of any node from $\Phi_i$. Remark that $f_i^i(r)=r$, $\forall i \in \mathcal{C}$.
\end{lemma}
\begin{IEEEproof}
	See Appendix~\ref{app:3}.
\end{IEEEproof}
Following the association policy defined later in Section \ref{Subsec:AssociationPolicy}, Lemma \ref{lemma:DistanceInterferer} defines a shield circle of radius $f_j^{i}(r)$ around the reference user that does not contain any interferer from the distribution $\Phi_i$. Thus, a logical consequence is that $f_j^{i}(r)$ an increasing function of the distance to the serving node $r$ from the distribution $\Phi_j$.

\subsection{Association Probabilities}

\subsubsection{Association Policy} \label{Subsec:AssociationPolicy}
We consider domination by the long-term statistics of the channel. The reference user connects to the UAV or the TBS providing the strongest average received power. Consequently, provided that the user is connected to a distribution $\Phi_i$ then the serving entity is the nearest node from $\Phi_i$ to the reference user.

\subsubsection{Association Probability to the Central UAV}
\label{Sec:AssociationCenter}
Let $A_c(r,Lu)$ denote the association probability to the central UAV given that it is in LOS condition with the user and distant by r. In contrast, $A_c(r,Nu)$ refers to the case of NLOS condition with the reference user.
\begin{lemma}[Central association probability]
	\label{lemma:Associationcenter}
	The expression of $A_c(r,i)$, $i$ $\in$ $\{Lu,Nu\}$, is given by:
\begin{align}
A_c(r,i)=   \prod_{j\in \mathcal{C}}^{} \exp\left(-2\; \pi\; \int_{0}^{f_i^j(r)} \lambda_j(x)\;x \;dx\right),
\end{align}

\end{lemma}
where $f_i^j(r)$ is given by Lemma~\ref{lemma:DistanceInterferer} and $\lambda_i$ is the density of the point process $\Phi_i$.
\begin{IEEEproof}
	See Appendix~\ref{app:4}.
\end{IEEEproof}

\subsubsection{Association Probability to a Non-Central UAV}
\label{Sec:AssociationNonCentral}
Let $A(r,i,\mathcal{S})$ denotes the association probability to the nearest neighbor, distant by $r \geq 0$, from the distribution $\Phi_i$ such that $\mathcal{S}=``r_c \leq R^*"$ refers to the case where the UAV is placed above the cluster center of the reference user and $\mathcal{S}=``r_c > R^*"$ refers to the case where the UAV is relocated according to Algorithm~\ref{algorithm}.
\begin{lemma}[Non-central association probability]
	\label{lemma:AssociationNonCentral}
	For $i \in \mathcal{C}$, the expression of $A(r,i,\mathcal{S})$ is given by : 
	\begin{align}
A(r,i,\mathcal{S})= \overline{A}(r,i,\mathcal{S}) \times \prod_{j\in \mathcal{C} \setminus i}^{} \exp\left(-2\; \pi\; \int_{0}^{f_i^j(r)} \lambda_j(x)\;x \;dx\right) ,
	\end{align}
	where $\overline{A}(r,i,\mathcal{S})$ is the probability that the serving entity from $\Phi_i$ is providing stronger signal than the central UAV. The expression of $\overline{A}(r,i,\mathcal{S})$ is given by : 
	\begin{align}
		\begin{split}
		&\overline{A}(r,i,\mathcal{S})=\int_{f_i^{\rm Lu}(r)}^{\infty} f_{r_{\rm cu}|\mathcal{S}}(x) \;\mathcal{P}_{\rm Lu}(x)\;dx + \int_{f_i^{\rm Nu}(r)}^{\infty} f_{\rm r_{\rm cu}|\mathcal{S}}(x) \;\mathcal{P}_{\rm Nu}(x)\;dx.
		\end{split}
	\end{align}
\end{lemma}
\begin{IEEEproof}
	See Appendix~\ref{app:5}.
\end{IEEEproof}

\subsection{Laplace Transform of Interference}
In this subsection, the Laplace transform of interference is derived. To this end, several scenarios are possible depending on what node the reference user is connected to.

\subsubsection{Reference user connected to the central UAV} Let  $\mathcal{L}_{I}^c(s,i,r)$ denotes the Laplace transform of the interference experienced at the reference user when it is connected to its central UAV distant by $r$ and conditioned on the placement strategy $\mathcal{S}$.
\begin{lemma}[Central Laplace interference]
	\label{lemma:InterferenceCentral}
	The Laplace transform of interference in the case where the UAV is connected to the central UAV, distant by $r\geq0$, conditioned on the channel type $i \in \{Lu,Nu\}$, is given by:
		\begin{figure*}[htp]
		\normalsize
		\begin{gather} 		
			\mathcal{L}_{I}^c(s,i,r)= \prod_{j\in \mathcal{C}}^{} \exp\left(-2\pi \int_{f_i^j(r)}^{\infty} \left(1-\left (  \frac{m_j}{m_j+s\rho_j \eta_j (x^2+h_j^2)^{-\frac{\alpha_j}{2}}}\right) ^{m_j} \right)\lambda_j(x)x dx\right). \label{eq:CentraLaplaceInterference}
		\end{gather}
		\vspace*{2pt}
	\end{figure*}%
\end{lemma}
The Laplace transform of the interference is used later to derive the coverage probability. It is a decreasing function of the distance to the serving node $r$, which means that the interference $I$ is increasing with $r$, by definition of the Laplace transform, i.e. $\mathcal{L}_{I}^c(s,i,r)= \mathbb{E} ( \exp\left(-s I\right))$ . This is a logical consequence since when the connected node is far then a large shield circle radius $f_j^i(r)$ is imposed around the reference user. 
\begin{IEEEproof}
	See Appendix~\ref{app:6}.
\end{IEEEproof}
\subsubsection{Reference user not-connected to the central UAV}
Suppose that the user is connected to the nearest point from the distribution $\Phi_i$, distant by $r_i$, and different from the central UAV. In the following lemma, we derive the Laplace transform of the interference generated by the central UAV. 
\begin{lemma}[Non-central Laplace interference]\label{lemma:InterferenceNoncentral}
The Laplace interference generated by the central UAV conditioned on the distance to the serving entity $r_i$ and  the placement strategy $\mathcal{S}$ is given by:
\begin{gather}
	\mathcal{L}_{I}(s,i,r,\mathcal{S})=\overline{\mathcal{L}_{I}}(s,i,r, \mathcal{S})   \times \prod_{j\in \mathcal{C}\setminus i}^{}\exp \left(-2 \pi \int_{f_i^{j}(r)}^{\infty}1- \left (  \frac{m_j}{m_j+s\rho_j \eta_j (x^2+h_j^2)^{-\frac{\alpha_j}{2}}}\right) ^{m_j} \lambda_{j}(x) x dx\right), 
\end{gather}
 where $\overline{\mathcal{L}_{I}}(s,r,i, \mathcal{S})$ is the Laplace transform of interference generated by the central UAV conditioned on the distance to the serving entity $r$ and  the placement strategy $\mathcal{S}$:
\begin{gather}
\overline{\mathcal{L}_{I}}(s,r,i, \mathcal{S})=\sum_{j \in \{ Lu, Nu\}}\int_{f_i^j(r)}^{\infty} \left (  \frac{m_j}{m_j+s \rho_j \eta_j (z^2+h^2)^{-\frac{\alpha_j}{2}}}\right) ^{m_j} \frac{ \mathcal{P}_j(z) f_{r_{\rm cu}|\mathcal{S}}(z)}{\int_{f_i^j(r_i)}^{\infty} f_{r_{\rm cu}| \mathcal{S}}(x)dx }dz.\label{Eq:LaplaceTransform3} 
\end{gather}

\end{lemma}
\begin{IEEEproof}
	See Appendix~\ref{app:7}.
\end{IEEEproof}

\subsection{Coverage probability}
In this section, we derive the analytical expression of the coverage probability $C(\gamma)$, defined as the probability that the received signal at the user is below a certain threshold $\gamma$. 
\begin{prop}
	\label{Proposition:Coverage}
 The coverage probability $C(\gamma)$ is given by:
	\begin{align}
		\begin{split}
		C(\gamma)= \int_{0}^{\infty} \Big(\sum\limits_{i \in \mathcal{C}} \mathds{E}_{\mathcal{S}}\{ L(r,i,\mathcal{S})A(r,i,\mathcal{S})\}  f_{R_i}(r) + \sum_{i \in \{\rm Lu,Nu\}}  L_c(r,i)A_c(r,i) \mathcal{P}_i(r) \mathds{E}_{\mathcal{S}}\{f_{r_{cu|S}}(r) \} \Big)  dr,
		\end{split}
\end{align}

	where the expectation over the placement strategy $\mathcal{S}$ are given by:
	
	\begin{gather}
		\label{Eq:Cov1}
		\mathds{E}_{\mathcal{S}}\{ L(r,i,\mathcal{S})A(r,i,\mathcal{S})\}=L(r,i,r_c\leq R^*)A(r,i,r_c \leq R^*) \left(1-\exp\left(-\lambda_{LBD} \pi {R^*}^2\right)\right) \\ \nonumber + \int_{R^*}^{\infty} L(r,i,r_c>R^*)A(r,i,r_c>R^*) f_{\rm R_c}(r_c)d r_c,  \\ 	\label{Eq:Cov2}   
		\mathds{E}_{\mathcal{S}}\{f_{r_{cu|S}}(r) \}= \int_{R^*}^{\infty} f_{r_{cu}|rc>R^*} (r) f_{R_c}(r_c) dr_c + \left(1- \exp\left(-\lambda_{LBD} \pi {R^*}^2\right)\right)f_{r_{cu}|rc\leq R^*} (r) .
	\end{gather}
Moreover, $L(r,i,\mathcal{S})$ is given by: 
	\begin{align}
		L(r,i,\mathcal{S})=\sum_{k=0}^{m_i-1} (-1)^k \frac{s^k}{k!} \frac{d^k}{ds^k} 	\mathcal{L}_{I+\sigma^2}(s_r,i,r,\mathcal{S}),
	\end{align}
	whereas $L_c(r,i)$ is given by: 
	\begin{align}
		L_c(r,i)=\sum_{k=0}^{m_i-1} (-1)^k \frac{s^k}{k!} \frac{d^k}{ds^k} 	\mathcal{L}_{I+\sigma^2}^c(s_r,i,r),
	\end{align}
	and finally $s_r$ is given by: 
	\begin{align}
		s_r=\frac{\gamma m_i (h_i^2+r^2)^{\alpha_i/2}}{ \rho_i \eta_i}.
	\end{align}
\end{prop}
\begin{IEEEproof}
	See Appendix~\ref{app:8}.
\end{IEEEproof}

\section{Numerical Results}
 \begin{table}[]\caption{System parameters~\cite{mypaper,Vertical}}
	\label{Table2}
	\centering
	\resizebox{0.85\textwidth}{!}{%
	\begin{tabular}{|c|c|c|c|}
		\hline
		Parameter                     & Value          & Parameter                            & Value \\ \hline
		\multicolumn{2}{|c|}{Spatial distribution}     & $\Delta f$                           &  1 Ghz     \\ \hline
		Simulation area               & 900 km$^2$               & $k$                                  & $5.92\times 10^6$ m$^{-1}$      \\ \hline
		$\lambda_{\rm LBD}$           & $0.52\times 10^{-6}$               & $\alpha$                             &   $10^{-6}$ m    \\ \hline
		$\lambda_{\rm cc}$            &$3\times 10^{-6}$                & $\delta_s$                           &$10^{-5}$       \\ \hline
		$\lambda_{\rm TBS}$           & $2\times 10^{-6}$               & $p_{\rm trans}$                      &     2 kW  \\ \hline
		\multicolumn{2}{|c|}{Environmental parameters} & $p_{\rm prop}$                             &  100 W     \\ \hline
		$\mathcal{B}$                 &  90\%              & $p_{comm}$                             &2 W       \\ \hline
		$a$                           &  1              & \multicolumn{2}{c|}{Other parameters}        \\ \hline
		$c$                           &  1              & $\eta_{\rm Lu}=\eta_{\rm Lb}$        & 0.9      \\ \hline
		$b_{\rm SubUrban}$            & 6.581               & $\eta_{\rm Nu}=\eta_{\rm Nb}$        & 0.7      \\ \hline
		$b_{\rm Urban}$               &0.151                & $\rho_{\rm Lu} = \rho_{\rm Nu}$      &   1 W    \\ \hline
		$b_{\rm denseUrban}$          &  0.101              & $\rho_{\rm Lb}=\rho_{\rm Nb}$        & 30 W      \\ \hline
		\multicolumn{2}{|c|}{Laser-link parameters}    & $\alpha_{\rm Lu}= \alpha_{\rm Lb}$   &  2     \\ \hline
		$D$                           &  0.1 m              & $ \alpha_{\rm Nu} = \alpha_{\rm Nb}$ & 3      \\ \hline
		$\Delta \theta$               &$3.4 \times 10^{-5}  $              &$h_{ \rm Lb}= h_{\rm Nb}$            &    25 m   \\ \hline
		$wT\chi$                      &  0.004 m$^2$              & $h_{ \rm Lu}= h_{\rm Nu}$        & 100 m      \\ \hline
		$C_n^2(n)$                    & $0.5  \times 10^{-14}$               & $m_{\rm Lu}=m_{\rm Lb}$             &1      \\ \hline
		$\hbar$                        &    $ 6.63\times 10^{-34} $ m$^2$  kg s$^{-1}$           & $m_{\rm Nu}= m_{\rm Nb}$           & 3      \\ \hline
	\end{tabular}
}
\end{table}
In this section, we validate the theoretical results we have proposed in this paper through Monte Carlo simulations. We analyze various figures to reveal some relevant trends related to the setup we have suggested. Finally, we provide high-level insights about the system we proposed for future practical deployment of laser-powered drones. In all of the figures, we use markers for the simulation results whereas solid lines are used to represent the analytic expressions.\par
\begin{figure}
	\centering
	\captionsetup{justification=centering}
	\includegraphics[width=3.5in]{"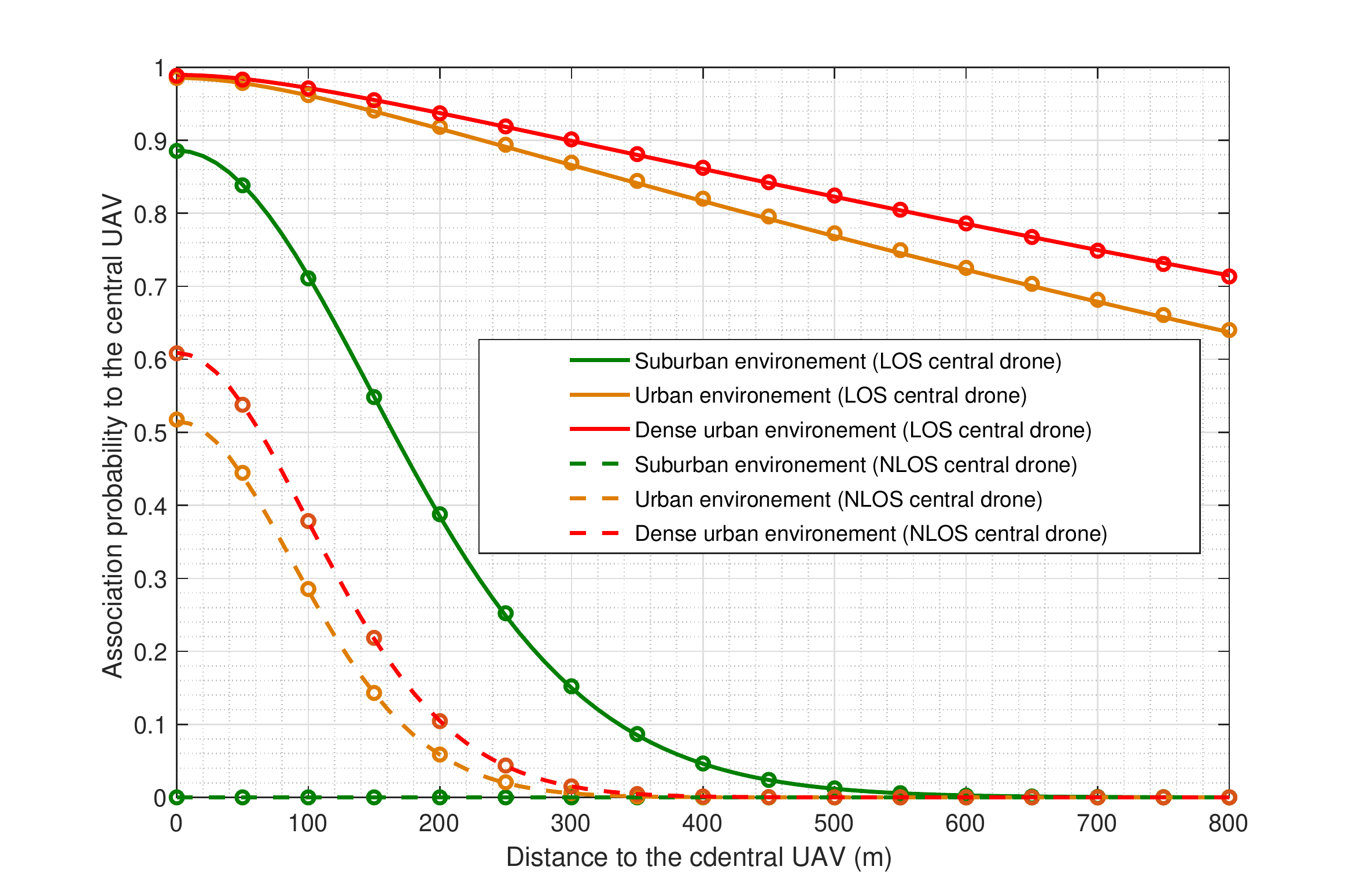"}
	\caption{Conditional association probability to the central UAV as a function of the serving distance.}
	\label{fig:AssociationProbability}
\end{figure}

In Figure~\ref{fig:AssociationProbability}, we plot the association probability conditioned on the distance to the reference user in different environments.  The figure shows that the probability to connect to the central UAV in dense urban environments is significantly higher than in suburban environments. This means that in dense urban environments, since links between drones and the reference user are often NLOS, as long as a LOS drone is placed close to the reference user there is a high chance to be the serving drone, this is not the case for suburban environments since links are probably LOS and the probability of having another LOS drone providing a stronger signal is not negligible. The same figure brings to light the effect of the link type on the association probability since the association to an NLOS central drone, plotted with dashed lines, is lower than the association probability to a central drone in LOS condition with the reference user, showing how much building blockage could affect the system setup.\par
\begin{figure}
	\centering
	\captionsetup{justification=centering}
	\includegraphics[width=3.5in]{"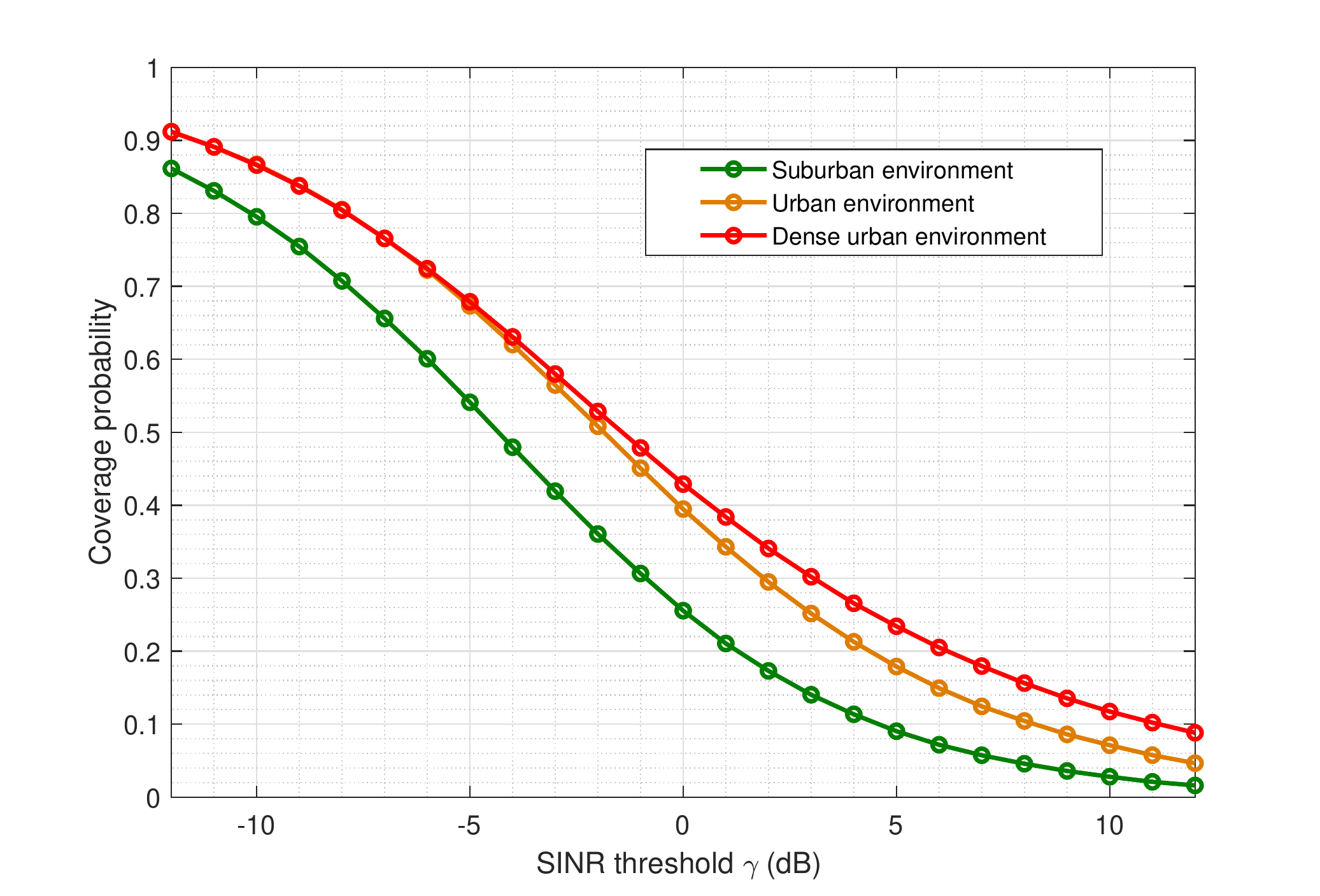"}
	\caption{Coverage Probability as a function of the SINR threshold.}
	\label{fig:CoverageSINR}
\end{figure}
\begin{figure*}
	\centering
	\begin{subfigure}{.5\textwidth}
		\centering
			\includegraphics[width=3.5in]{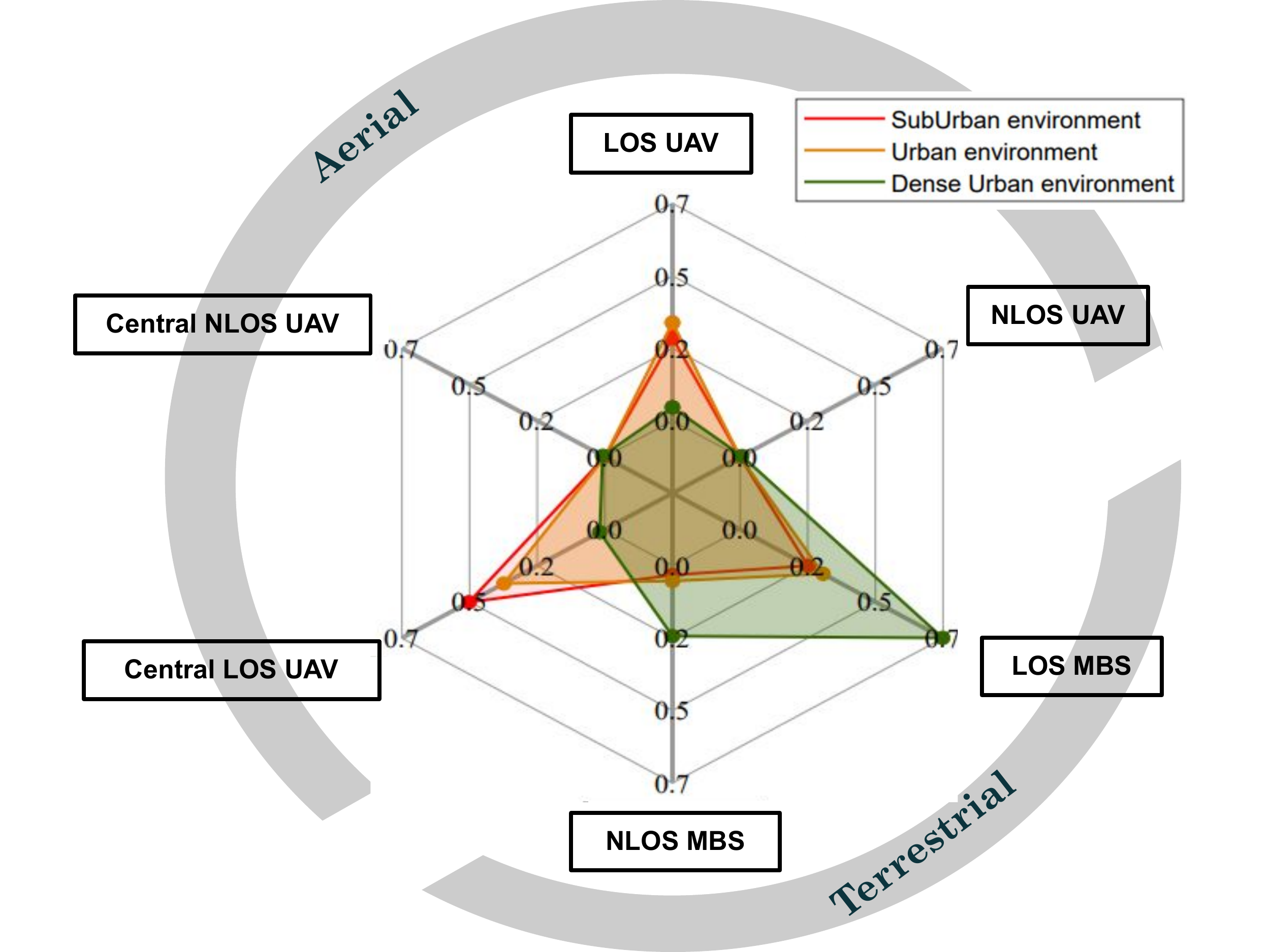}
	\caption{Environment effect on the connectivity profile.}
		\label{fig:sub1}
	\end{subfigure}%
	\begin{subfigure}{.5\textwidth}
		\centering
		\includegraphics[width=3.5in]{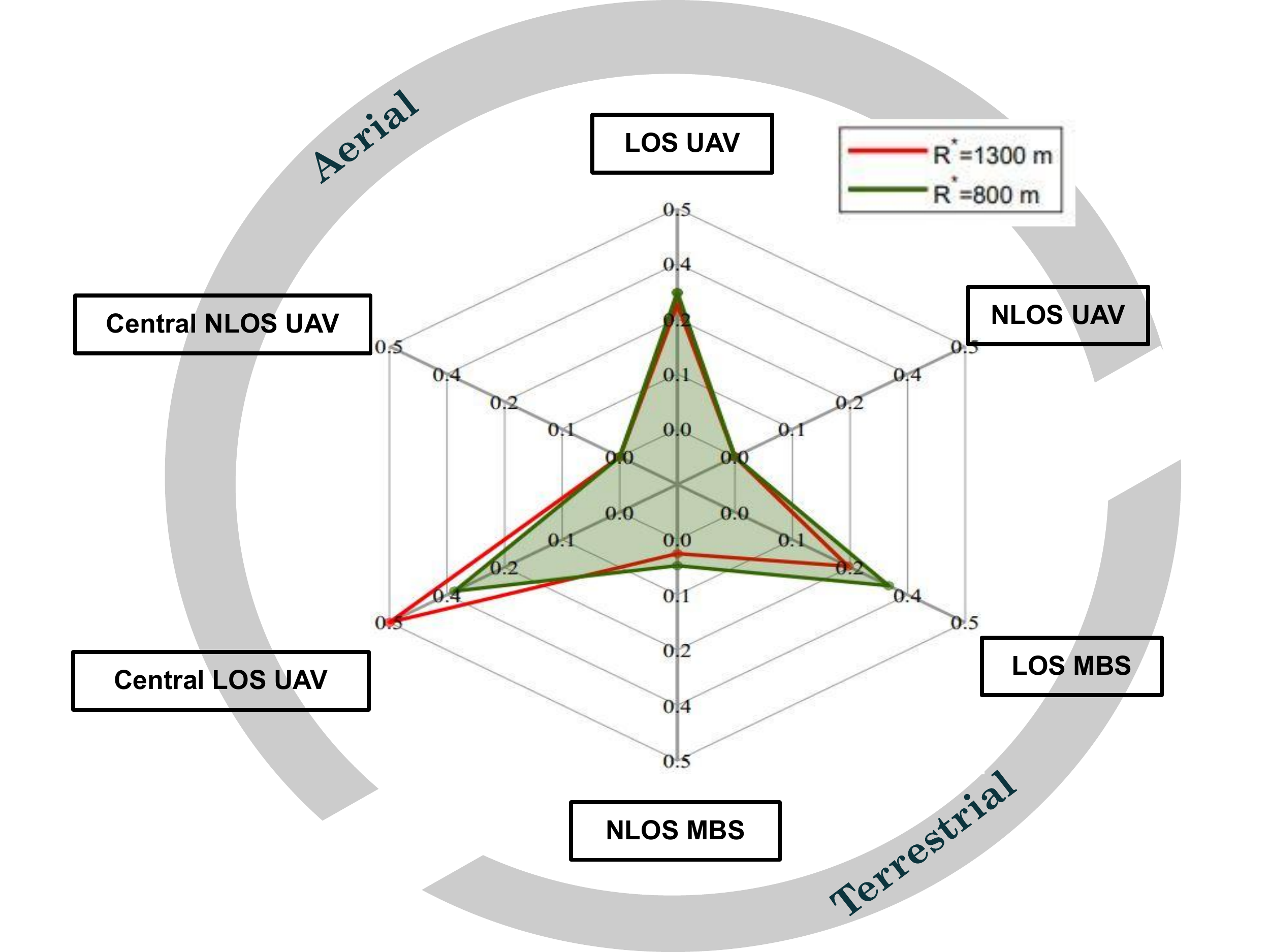}
			\caption{ Effect of the critical charging distance $R^*$.}
		\label{fig:sub2}
	\end{subfigure}
	\caption{Average user connectivity profile.}
	\label{fig:Co}
\end{figure*}

Figure~\ref{fig:CoverageSINR} shows the coverage probability as a function of the signal-to-interference-plus-noise ratio (SINR) threshold $\gamma$ in different type of environments. It shows that coverage in dense urban environments is higher than coverage in  urban  and suburban environments. The degradation of wireless coverage is a logical consequence of the significant interference generated by all the nodes in the system, especially when most of them are in LOS condition with the user. In other words, it is a confirmation of Figure~\ref{fig:Co}, but from a different angle.\par 
\begin{figure}
	\centering
	\captionsetup{justification=centering}
	\includegraphics[width=3.5in]{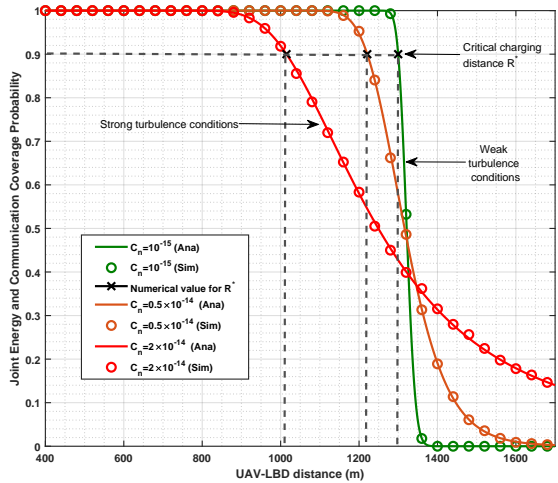}
	\caption{Joint energy and communication coverage as function of the distance to the drone.}
	\label{fig:Lasercharging}
\end{figure}
Figure~\ref{fig:Lasercharging} shows the joint energy and communication coverage probability as a function of the distance to the drone R under different turbulence conditions. The figure is informative regarding the charging requirement for the drones as the coverage probability decreases when increasing the critical charging distance, in other words in strong turbulent environments. Moreover, we can state that reasonable ranges for the critical charging distance are [1000 1300] meters for energy and communication coverage threshold  $\mathcal{B}=90\%$ . The critical charging distance is also related to the energy coverage and communication threshold $\mathcal{B}$, since it decreases when targeting highly reliable charging and backhaul link for the drones when increasing $\mathcal{B}$. 

Figure~\ref{fig:Co} shows the connectivity diagram of the user in different types of environments and in different laser charging coverage. The purpose of this figure is to answer the question to what extent the UAV network could support the terrestrial architecture. We distinguish 3 different connection modes: (1) connection to the central UAV, (2) connection to any other UAV, and (3) connection to a TBS. As we consider LOS and NLOS links, the number of connection modes increases to 6 accordingly. The trends in Figure~\ref{fig:sub1} show a considerable tendency to connect to terrestrial architecture in dense urban environments. This is due to the fact that LOS links to UAVs are largely affected by building blockage. In addition, the transmit power of UAVs (1W) is remarkably lower than the transmit power of TBSs (30W), which explains why links to UAVs are very sensitive to LOS probability. That said, in urban and suburban environments, the drone network successfully assists the TBS by providing stronger signals to users. We note that the connection rate to airborne base stations is 7.4\% for in dense urban environments, 66.4\% in urban environments and increases to 73.5\% in suburban environments.
In a nutshell, this confirms that UAVs must be deployed in LOS conditions to be effective in wireless coverage scenarios, i.e. in urban to suburban environments. Figure~\ref{fig:sub2} captures the effect of the maximal laser charging radius on the user connectivity. We can easily notice that under poor laser charging conditions, i.e under low critical charging distance $R^*=800 m$, the association rate to the central UAV is 33.8\%. However, under better laser critical charging distance,  $R^*=1300 m$, the association probability to the central drone increases to 47.16\%. This is the consequence of expanding the drones' flexibility, allowing them to be positioned close to user clusters. At the same time, the association rate to the TBSs decreases by 11\% when increasing the critical charging distance, as the performance of the drones is severely affected.

\begin{figure}
	\centering
	\captionsetup{justification=centering}
	\includegraphics[width=3.5in]{"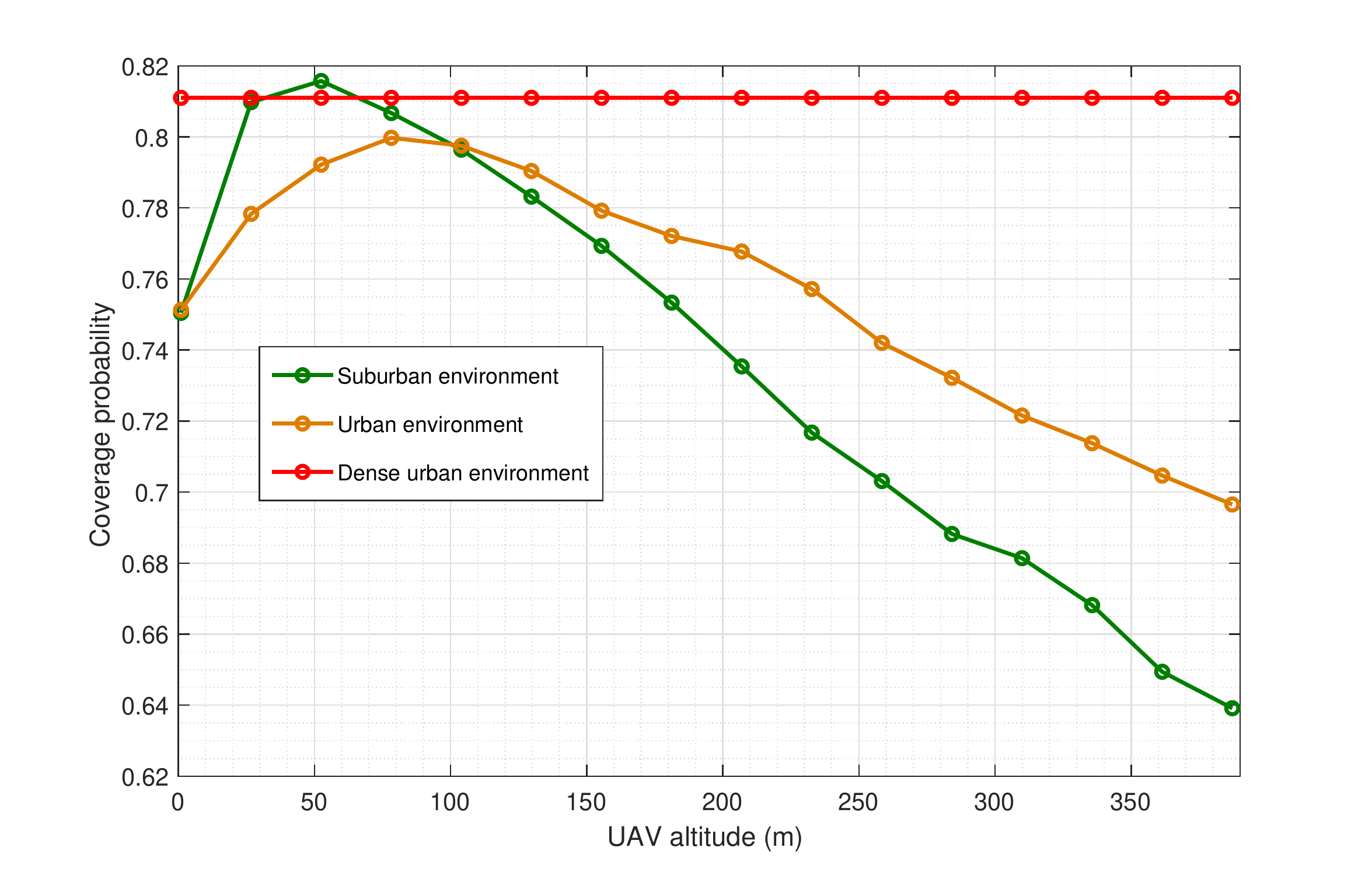"}
	\caption{UAV altitude effect on the coverage probability.}
	\label{fig:CoverageAltitude}
\end{figure}
In Figure \ref{fig:CoverageAltitude}, we plot the coverage probability as a function of drone altitude for different building densities. This figure shows how drone altitudes can affect the users' quality of service in certain scenarios. With the current system configuration, we can see that the optimal drone placement is at 50 m altitude in a suburban environment and at 80 m altitude in an urban environment. The same figure confirms the results observed in Figure \ref{fig:Co} where the coverage probability is not affected by the altitudes of the UAVs in dense urban environments because the probability of association to the aerial network is low. This problem can be solved by optimizing the altitude of each drone individually, which is not feasible in our system model. \par
	\begin{figure}
	\centering
	\captionsetup{justification=centering}
	\includegraphics[width=3.5in]{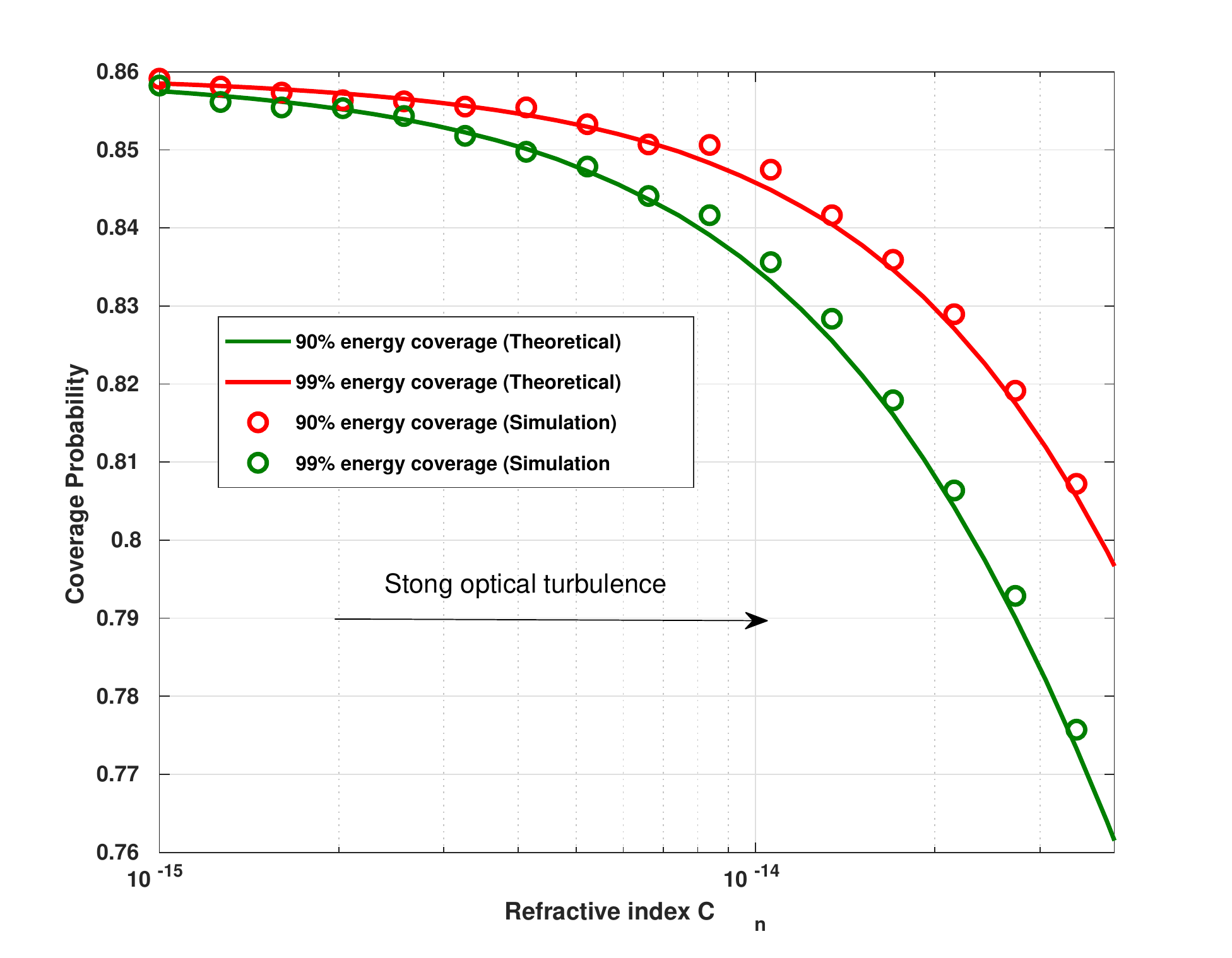}
	\caption{Optical turbulence effect on the coverage probability for different energy coverage ($\mathcal{B}$) levels.}
	\label{fig:RefractiveIndex}
\end{figure}
Figure \ref{fig:RefractiveIndex} presents the turbulence effect on the coverage probability for different levels of the drone energy reliability, denoted by $\mathcal{B}$. This figure is equivalent to plotting against the critical charging distance $R^*$ since there is a direct relation between $R^*$ and $\mathcal{B}$. The figure shows a degradation in the coverage probability for moderate-to-strong optical turbulence and relatively constant energy coverage probability for weak optical turbulence ($C_n^2=10^{-15}$). We can also remark that coverage probability at the user is affected by the energy coverage reliability level. For instance, the coverage probability is better for low joint energy and communication coverage levels. We can also notice that the gap induced by the drone energy coverage increases in moderate to strong optical turbulence. The aforementioned observations could be explained as follows; (1) when increasing the energy coverage threshold $\mathcal{B}$ the critical charging distance $R^*$ decreases to meet this requirement.  Consequently, this fact induces shifting the UAV more close to their nearest LBDs instead of being positioned above the cluster center, thus wireless coverage is sacrificed for better energy coverage. (2) The same phenomenon happens when increasing the refractive index, i.e increasing the optical turbulence effect, as the critical charging distance decreases in such environments. In a nutshell, a trade-off between energy and wireless coverage is visible in all the figures we have presented as a logical consequence of the placement strategy we have proposed. The best-case scenario corresponds to the placement of the drone above the cluster centers while maintaining energy harvesting, which yields the best average performance. However, in the worst case, drones are attracted to the energy sources, and therefore users tend to connect to the available terrestrial network architecture. 
\section{Conclusion}
In this paper, we have presented a comprehensive coverage analysis for assisting a terrestrial network architecture through a laser-powered drone network. We used stochastic geometry to model the system and proposed a novel deployment strategy for the laser-powered UAVs that ensures concomitantly energy and wireless coverage for GUEs. As proposed in our previous work,  we assume simultaneous communication and energy harvesting at the drone, enabling a reliable backhaul link. Within this framework, we shed light on an interesting trade-off between energy reliability and coverage efficiency. We have also showed how optical turbulence badly affect the drone's wireless charging and consequently results in a degradation of the user's coverage probability. In the light of these analyses, we believe that laser-powered UAVs strongly compete with tethered, solar-powered, and untethered UAVs and could offer interesting alternatives for future wireless communication networks. 
\vspace{-4mm}
\appendix
\vspace{-1mm}

\vspace{-4mm}
\subsection{Proof of Lemma~\ref{lemma: CriticalChargingDistance}}\label{app:1}
To start with, we have previously demonstrated in~\cite{mypaper} that the joint energy and communication coverage probability is dominated by the energy coverage probability, and thus to derive the critical charging distance, we rely on lower bounding the energy coverage probability defined in the same work.
Let us denote the lower bound on the energy coverage probability by $\mathcal{B} \in [0\; 1]$:
\begin{align}
	P_{\rm energy}&=\mathbb{P}\left(h_t p_{\rm harv}(r) \geq p_{prop}+p_{comm} \right) \geq \mathcal{B}. \notag \\
	\Rightarrow&\ \mathbb{P}(h_t  \geq \frac{p_{\rm prop}+p_{\rm comm}}{p_{\rm harv}(r)} ) \geq \mathcal{B}.\notag\\ 
	\Rightarrow&\ 1-F_{h_t}(\frac{p_{\rm prop}+p_{\rm comm}}{p_{\rm harv}(r)})\geq \mathcal{B}. \notag\\ 
	\overset{(a)}{\Rightarrow}&\ \frac{1}{2}-\frac{1}{2}\erf\left(\frac{\ln(\frac{p_{\rm prop}+p_{\rm comm}}{p_{\rm harv}(r)}) + 2\sigma}{\sqrt{2} \sigma} \right) \geq \mathcal{B}.\notag \\ \label{eq:Lemma1}
	\overset{(b)}{\Rightarrow}&\	\frac{\ln(\frac{p_{\rm prop}+p_{\rm comm}}{p_{\rm harv}(r)}) + 2\sigma}{\sqrt{2} \sigma} \leq \erf^{-1}\left( 1 -2 \mathcal{B}\right).
\end{align}
where (a) is due to the fact that the turbulence effect is following a lognormal distribution and (b) is due to the fact that $\erf ^{-1}$ is an increasing function. Based on the inequality~(\ref{eq:Lemma1}), we conclude that $R^*$ should satisfy (\ref{eq:Rstar})  provided in Lemma~\ref{lemma: CriticalChargingDistance}.
\vspace{-3mm}
\subsection{Proof of Proposition~\ref{Proposition:PDFrcu}}\label{app:2}
To derive the PDF of the distance between the reference user and the central UAV, denoted by $r_{cu}$, in the case of $ \mathcal{S}=``r_c > R^*"$, we start by deriving $F_{r_{cu}| r_c> R^* }$. Let $\alpha \in \mathbb{R}_+$, we can write the following:
\begin{align}
	F_{r_{cu}| r_c> R^* }(\alpha)&=	\mathbb{P}(r_{\rm cu} \leq \alpha |r_c > R^{* } )
	= \mathbb{P}(r_{\rm cu}^2 \leq \alpha^2 |r_c > R^{* } ).
	\end{align}
We exploit the cosine rule applied to the colored triangle shown is Figure \ref{fig:SystemModelSummary}. Notice that $r$, $d$, and $\theta$ are random variables, but, since we conditioned on $r_c$ then $r_{cc}$ is also fixed due the fact that $r_{cc}=r_c-R^*$. 
	 \begin{align}
	 &=\mathbb{E}_{\rm r}\left( \mathbb{P}\left(r_{\rm cc}^2 +r^2 - 2 r r_{\rm cc}    cos(\theta ) \leq \alpha^2 \; \Big| \; r_c> R^*,r\right)\right)\nonumber, \\ &=\mathbb{E}_{r}\left(\mathbb{P}\left( \cos(\theta)\geq \frac{r_{\rm cc}^2 +r^2 - \alpha^2}{2 r r_{\rm cc} }\; \Big| \;r_c> R^*, r\right)\right)
	\nonumber, \\&= \mathbb{E}_{\rm r} \left(\mathbb{P}\left( \cos(\theta)\geq \frac{r_{\rm cc}^2 +r^2 - \alpha^2}{2 r r_{\rm cc} }\big| r_c> R^*,r\right)\right ). 
		\end{align}
	Now using the fact that $\arccos$ is a decreasing function defined on $[-1 \; \;1]$ and that $\theta$ is uniformly distributed across $[0 \; \; 2\pi]$, we can further develop as follows: 
	\begin{align}
	&=\mathbb{E}_{\rm r} \Bigg(\frac{ \arcos \left(\frac{r_{\rm cc}^2+r^2-\alpha^2}{2rr_{\rm cc}}\right) }{\pi}\mathds{1}\Big\{\Big|  \frac{r_{\rm cc}^2+r^2-\alpha^2}{2r r_{\rm cc}}\Big| \leq 1\Big\}+ \mathds{1}\Big\{\frac{r_{\rm cc}^2+r^2-\alpha^2}{2r r_{\rm cc}} < -1\Big\} \nonumber \\&+ 0 \times \mathds{1}\Big\{\frac{r_{\rm cc}^2+r^2-\alpha^2}{2r r_{\rm cc}} > 1\Big\}\Bigg),
	\nonumber \\&\stackrel{(a)}{=}\mathbb{E}_{\rm r} \left(   \frac{  \arcos\left(\frac{r_{\rm cc}^2+r^2-\alpha^2}{2r r_{\rm cc}}\right) }{\pi  }\mathds{1}\Big\{| \alpha-r_{\rm cc}|\leq r\leq r_{\rm cc}+\alpha\Big \}\right)  + \mathbb{E}_{r} \Bigg( \mathds{1}\Big\{r< \alpha -r_{\rm cc} \Big\}\Bigg).
\end{align}
At this stage, we average over the random variable $r$ taking into account that r is derived form the distance of one point and the cluster center from a MCP. Thus we can write the following: 
	\begin{align}
	&= \int_{[0 , r_{max}] \cap [| \alpha-r_{\rm cc}| ,  r_{\rm cc}+\alpha ]}\frac{2r}{\pi r_{max}^2} \arcos\left(\frac{r_{\rm cc}^2+r^2-\alpha^2}{2 r_{\rm cc} r}\right)dr+ \frac{{(\alpha -r_{\rm cc})}^2}{r_{max}^2}\mathds{1}\Big\{0 \leq \alpha -r_{\rm cc}\leq r_{max}\Big\}.
\end{align}
In what follows, we explain how (a) is derived. Notice that $r$, $r_{cc}$, and $\alpha$ are positive:
\begin{align}
	&\mathds{1}\Big\{\Big|  \frac{r_{\rm cc}^2+r^2-\alpha^2}{2r r_{\rm cc}}\Big| \leq 1\Big\} = \mathds{1}\Big\{ r_{\rm cc}^2+r^2-\alpha^2 \leq  2r r_{\rm cc}  \Big\}\times\nonumber \mathds{1}\Big\{ r_{\rm cc}^2+r^2-\alpha^2 \geq  -2r r_{\rm cc}  \Big\}, \nonumber \\&=\mathds{1}\Big\{ (r_{\rm cc}-r)^2 \leq \alpha^2   \Big\} \mathds{1}\Big\{ (r_{\rm cc}+r)^2 \geq \alpha^2   \Big\}\nonumber,  \\&=\mathds{1}\Big\{ |r_{\rm cc}-r| \leq \alpha   \Big\}\mathds{1}\Big\{ |r_{\rm cc}+r| \geq \alpha   \Big\}\nonumber, \\&=\mathds{1}\Big\{ r_{\rm cc}-\alpha \leq r \leq r_{\rm cc}+\alpha   \Big\}\mathds{1}\Big\{ r \geq \alpha-r_{\rm cc}   \Big\}\nonumber, \\&=\mathds{1}\Big\{ |\alpha -r_{\rm cc}| \leq r \leq r_{\rm cc}+\alpha   \Big\}.
\end{align}
Now that we derived the CDF of $r_{cu}$, we obtain the PDF expression by deriving the CDF with respect to $\alpha$, to this end, we simply apply the Leibniz integral rule to the CDF to get the expression described in Proposition~\ref{Proposition:PDFrcu}. Different cases should be distinguished depending on the result of $[0 , r_{max}] \cap [| \alpha-r_{\rm cc}| ,  r_{\rm cc}+\alpha ]$.

\subsection{Proof of Lemma~\ref{lemma:DistanceInterferer}}\label{app:3}
Let $r_i$ be the distance from  the reference user to its nearest neighbor from a point process $\Phi_i$. Suppose that the reference user is connected to the nearest neighbor from the point process $\Phi_j$, distant by $r_j$. According to the association policy that implies connection to the strongest signal, the nearest interferer from $\Phi_i$ to the reference user must satisfy the following equation: 
\begin{gather}
	\rho_j \; \eta_j \; (h_j^2 + r_j^2)^{-\frac{\alpha_j}{2}} \geq  \rho_i \; \eta_i\; (h_i^2 + r_i^2)^{-\frac{\alpha_i}{2}} \nonumber, \\
	\left( \frac{\rho_i \; \eta_i\;}{\rho_j\; \eta_j} \right)^{\frac{2}{\alpha_i}}    \; (h_j^2 + r_j^2)^{\frac{\alpha_j}{\alpha_i}}-h_i^2 \leq   r_i^2\nonumber, \\
	\sqrt{\max\left(0,\left( \frac{\rho_i \; \eta_i\;}{\rho_j\; \eta_j} \right)^{\frac{2}{\alpha_i}}    \; (h_j^2 + r_j^2)^{\frac{\alpha_j}{\alpha_i}}-h_i^2 \right)}\leq   r_i.
\end{gather}
Hence, conditioned on the distance to the serving node $r$,  the nearest interferer distance is provided by: \begin{align}\begin{split} f_j^{i}(r)= \sqrt{\max\left( 0, (\frac{\rho_i \eta_i}{\rho_j \eta_j})^{\frac{2}{\alpha_i}}(h_j^2+r^2)^\frac{\alpha_j}{\alpha_i} -h_i^2\right)}, i,j \in \{\rm Lu,Nu,Lb,Nb\}.\end{split}
\end{align}

\subsection{Proof of Lemma~\ref{lemma:Associationcenter}}\label{app:4}
Suppose that the central UAV is located at a distance $r \geq 0$ with channel type $i$, where $i$ could be either LOS $(i=Lu)$, or NLOS $(i=Nu)$. Then, based on the association policy defined in Sec.~\ref{Subsec:AssociationPolicy}, the association probability to the central UAV is given by: 
\begin{align}
\nonumber 	&A_c(r,i) = \mathbb{P}\left( \overline{\mathcal{R}}_i(r) \geq \max \limits_{j\in \mathcal{C}} \left(\overline{\mathcal{R}}_j(r_j) \right)  \right),\\\nonumber 
	 &= \prod_{j\in \mathcal{C }}\mathbb{P}\Bigg( \overline{\mathcal{R}}_i(r) \geq \overline{\mathcal{R}}_j(r_j)   \Bigg),\\
	&=\prod_{j\in\mathcal{C}}\mathbb{P}\Bigg( r_j \geq f_i^j(r)\Bigg).
\end{align}
Since $r_j$ is the nearest neighbor distance from the PPP $\Phi_j$ with density $\lambda_j$, we can write the following: 
\begin{align}
A_c(r,i)=   \prod_{j\in\mathcal{C}}^{} \exp\left(-2\; \pi\; \int_{0}^{f_i^j(r)} \lambda_j(x)\;x \;dx\right).
\end{align}

\subsection{Proof of Lemma~\ref{lemma:AssociationNonCentral}}\label{app:5}
Let $A(r,i,\mathcal{S})$ denotes the association probability to the nearest neighbor, distant by $r \geq 0$, from the distribution $\Phi_i$, and $\mathcal{S}$ =\{ ``$r_c \leq R^*$",``$r_c > R^*$" \} denotes the deployment strategy specified in Algorithm~\ref{algorithm}. Accordingly, for $i \in \{Lu,Nu,Lb,Nb \}$,  we can write the following: 
\begin{align}
	&A(r,i,\mathcal{S}) =\overline{A}(r,i,\mathcal{S})  \mathbb{P}\Bigg(\overline{\mathcal{R}}_i(r) \geq \max \limits_{j\in \mathcal{C} \setminus i}\nonumber  \Big(\overline{\mathcal{R}}_j(r_j) \Big)  \Bigg),\\ \nonumber 
	& =\overline{A}(r,i,\mathcal{S}) \prod_{j\in \mathcal{C} \setminus i}\mathbb{P}\Bigg( \overline{\mathcal{R}}_i(r)\geq  \overline{\mathcal{R}}_j(r_j)   \Bigg),\\\nonumber 
	&=\overline{A}(r,i,\mathcal{S})\prod_{j\in \mathcal{C}\setminus i}\mathbb{P}\Bigg( r_j \geq f_i^j(r)\Bigg),\\
	=& \overline{A}(r,i,\mathcal{S})  \prod_{j\in \mathcal{C}\setminus i}^{} \exp \left(-2 \pi \int_{0}^{f_i^j(r)} \lambda_j(x)x \;dx \right),
\end{align}
where $\overline{A}(r,i,\mathcal{S})$  is the probability that the signal from the serving entity from $\Phi_i$ is stronger than the signal from the central UAV, given that the serving point is at a distance r from the reference user and conditioned on the placement of the central UAV. Let $r_{cu}$ be the distance from the reference user to the central UAV. The expression $\overline{A}(r,i,\mathcal{S})$ is given by:
\begin{align}
 & \overline{A}(r,i,\mathcal{S})=\mathbb{P}\Bigg( \overline{\mathcal{R}}_{\rm Lu}(r_{\rm cu}) \leq \overline{\mathcal{R}}_{\rm i}(r_{\rm i})  , LOS\Bigg)+ \nonumber \mathbb{P}\Bigg( \ \overline{\mathcal{R}}_{\rm Nu}(r_{\rm cu}) \leq  \overline{\mathcal{R}}_{\rm i}(r_{\rm i})  , NLOS \Bigg)\nonumber,  \\
	&=  \mathbb{P}\Bigg( r_{\rm cu} \geq f_i^{\rm Lu}(r),LOS \Bigg)+\mathbb{P}\Bigg( r_{\rm cu} \geq f_i^{\rm Nu}(r),NLOS \Bigg)\nonumber, \\
	&=\int_{f_i^{\rm Lu}(r)}^{\infty} f_{r_{\rm cu}|\mathcal{S}}(r) \;\mathcal{P}_{\rm Lu}(r)\;dr +\int_{f_i^{\rm Nu}(r)}^{\infty} f_{r_{\rm cu}|\mathcal{S}}(r) \;\mathcal{P}_{\rm Nu}(r)\;dr.
\end{align}

\subsection{Proof of Lemma~\ref{lemma:InterferenceCentral}}\label{app:6}

Suppose that the reference user is connected to its central UAV. Let $I_{i}$, $i \in \mathcal{C}$ denotes the interference induced by the point process $\Phi_i$. Let $I_c$ denotes the interference generated by the central UAV. Since the UAV is connected the central UAV placed at $x_c$, $I_c =0$, and hence, $\mathcal{L}_{I_c}(s)=1$. Let 	$\mathcal{L}^c_{I}(s,i,r)$ denotes the Laplace interference given that the reference user is connected to the central UAV with channel type $i \in \{Lu, Nu\}$, and $I$ denotes the total interference:
\begin{align}
	I&=\sum_{j \in\mathcal{C}} I_j\nonumber,  \\
	I&=\sum_{j \in \mathcal{C}} \sum_{x_i \in \Phi_j \setminus x_c}\mathcal{R}_{\rm j}(r(\ x_i)).
\end{align}
Hence, the Laplace transform of interference is given by:
\begin{align}
	&\mathcal{L}_{I}^c(s,i,r)= \mathbb{E} ( \exp\left(-s I\right))\nonumber, \\
&= \prod_{j\in \mathcal{C}}^{}\mathbb{E}\left(\exp\left(-s \sum_{x_i \in \Phi_j \setminus x_c}\mathcal{R}_{j}(r(x_i))\right)\right)\nonumber, \\
	&\stackrel{(a)}{=} \prod_{j\in \mathcal{C}}^{} \exp \left(-2 \pi \int_{f_i^j(r)}^{\infty} \left(1- \mathbb{E}_{G_j} \left \{ \exp\left(-s\mathcal{R}_{j}(x)\right)\right\} \right) \lambda_j(x) x dx\right)\nonumber, \\
	&\stackrel{(b)}{=} \prod_{j\in \mathcal{C}}^{} \exp\left(-2 \pi \int_{f_i^j(r)}^{\infty} \left(1-\left(  \frac{m_j}{m_j+s\overline{\mathcal{R}}_{\rm j}(x)}\right) ^{m_j} \right)\lambda_j(x) x dx\right),
\end{align}
where (a) is performed by applying the probability generating functional of a process~\cite{haenggi} and (b) is due to the fact that the $G_j$ is gamma distributed with channel-dependent fading parameters $m_j$.
\subsection{Proof of Lemma~\ref{lemma:InterferenceNoncentral}}\label{app:7}

Suppose that the UAV is not  connected to the central UAV, placed at $x_c$, with channel type $j$, instead it is connected to the nearest point from the distribution $\Phi_i$, distant by $r_i$, and placed at $x_i$. Suppose also that the placement strategy $\mathcal{S}$ is used. The generated interference expression is given by:
\begin{align}
	I&=I_c + \sum_{j \in \mathcal{C}\setminus i}  I_j,  \\
\end{align}
where $I_c$ is the interference generated by the central UAV, and $I_j$ is the interference generated by the other point porcesses except the point process $\Phi_i$ since $I_i=0$. Hence, based on the proof in Appendix.~\ref{app:5}, we can write the Laplace transform of interference as follows:
\begin{align}
	&\mathcal{L}_{I}(s,r,i, \mathcal{S})=\overline{\mathcal{L}_{I}}(s,r,i, \mathcal{S}) \times \nonumber \\&\prod_{j\in \mathcal{C} \setminus i}^{}\exp \left(-2 \pi \int_{f_i^{j}(r)}^{\infty}1- \Big (  \frac{m_j}{m_j+s\overline{\mathcal{R}}_{\rm j}(x) }\Big) ^{m_j} \lambda_{j}(x) x dx\right),
\end{align}
where the Laplace transform of the interference generated by the central UAV,   $\overline{\mathcal{L}_{I}}(s,r,i, \mathcal{S}) $,   is given by:
\begin{align}
	&\overline{\mathcal{L}_{I}}(s,r,i, \mathcal{S})=\mathbb{E}(exp(-s \; I_c))\nonumber, \\
	&=\mathbb{E}_{r(x_c)}\Big( \mathbb{E}_{j,G_j}(exp (-s \;\mathcal{R}_{\rm j}(r(x_{\rm c}) ) )\Big) \nonumber, \\
	&\stackrel{(a)}{=}\sum_{j \in \{ Lu, Nu\}} \int \Big (  \frac{m_j}{m_j+s \overline{\mathcal{R}}_{\rm j}(z) }\Big) ^{m_j} \mathbb{P}\left(r(x_c)=z, j| \mathcal{S}, \overline{\mathcal{R}}_{\rm j}(r(x_{\rm c}))\leq \overline{\mathcal{R}}_{\rm i}(r(x_{\rm i}))\right) dz\nonumber, \\
	&= \sum_{j \in \{ Lu, Nu\}} \int \Big (  \frac{m_j}{m_j+s\overline{\mathcal{R}}_{\rm j}(z)}\Big) ^{m_j} \frac{\mathbb{P}(\overline{\mathcal{R}}_{\rm j}(r(x_{\rm c}))\leq \overline{\mathcal{R}}_{\rm i}(r(x_{\rm i})),r(x_c)=z, j| \mathcal{S})}{ \mathbb{P}( \overline{\mathcal{R}}_{\rm j}(r(x_{\rm c}))\leq \overline{\mathcal{R}}_{\rm i}(r(x_{\rm i}))| \mathcal{S})} dz\nonumber, \\
	&= \sum_{j \in \{ Lu, Nu\}} \int \Big (  \frac{m_j}{m_j+s\overline{\mathcal{R}}_{\rm j}(z)}\Big) ^{m_j} \frac{\mathbb{P}(\overline{\mathcal{R}}_{\rm j}(r(x_{\rm c}))\leq \overline{\mathcal{R}}_{\rm i}(r(x_{\rm i})),r(x_c)=z, j| \mathcal{S})}{ \mathbb{P}( \overline{\mathcal{R}}_{\rm j}(r(x_{\rm c}))\leq \overline{\mathcal{R}}_{\rm i}(r(x_{\rm i}))| \mathcal{S}) }dz\nonumber, \\
	&\stackrel{(b)}{=}\sum_{j \in \{ Lu, Nu\}} \int_{f_i^j(r)}^{\infty} \Big (  \frac{m_j}{m_j+s \overline{\mathcal{R}}_{\rm j}(z)}\Big) ^{m_j} \frac{ \mathcal{P}_{\rm j}(z) f_{r_{\rm cu}|\mathcal{S}}(z)}{\int_{f_i^j(r)}^{\infty} f_{r_{\rm cu}| \mathcal{S}}(x)dx }dz.
\end{align}
(a) is due to the moment generating function of Gamma distribution. To explain (b), we can write:
\begin{align}
	&\mathbb{P}\Big( \overline{\mathcal{R}}_{\rm j}(r(x_{\rm c}))\leq \overline{\mathcal{R}}_{\rm i}(r(x_{\rm i}))| \mathcal{S}\Big) =	\int_{f_i^j(r)}^{\infty} f_{r_{\rm cu}| \mathcal{S}}(x)dx\nonumber, \\
	&\mathbb{P}\Big(\overline{\mathcal{R}}_{\rm j}(r(x_{\rm c}))\leq \overline{\mathcal{R}}_{\rm i}(r(x_{\rm i})),r(x_c)=r, j| \mathcal{S}\Big)=\nonumber \mathbb{P}\Big(\overline{\mathcal{R}}_{\rm j}(r(x_{\rm c}))\leq \overline{\mathcal{R}}_{\rm i}(r(x_{\rm i}))|r(x_c)=r, j, \mathcal{S}\Big)  \mathbb{P}\Big(r(x_c)=r, j|\mathcal{S}\Big)\nonumber, \\
	&=\mathbb{P}\Big(\overline{\mathcal{R}}_{\rm j}(r(x_{\rm c}))\leq \overline{\mathcal{R}}_{\rm i}(r(x_{\rm i}))|r(x_c)=r, j, \mathcal{S}\Big)\mathbb{P}\Big(j|r(x_c)=r,\mathcal{S}\Big)\times \nonumber \mathbb{P}\Big(r(x_c)=r|\mathcal{S}\Big)\nonumber, \\
	&= \mathds{1}\big( r \geq f_i^j(r) \big) \mathcal{P}_j(r) f_{r_{\rm cu}|\mathcal{S}}(r).
\end{align}

\subsection{Proof of Proposition~\ref{Proposition:Coverage}}\label{app:8}
\begin{align}
	&C(\gamma)= \mathbb{P}(\frac{S}{I+\sigma^2} \geq \gamma)\nonumber \\&=\mathbb{E}_{\mathcal{S},r,i|r,I|r,i,\mathcal{S}}  \Big\{  \mathbb{P}(\frac{\rho_i \eta_i G_i (h_i^2+r^2)^{-\frac{\alpha_i}{2}}}{I+\sigma^2} \geq \gamma)\Big\}\nonumber , \\
	&=\mathbb{E}_{\mathcal{S},r,i|r,I|r,i,\mathcal{S}}  \Big\{  \mathbb{P}(G_i\geq \frac{ \gamma (I+\sigma^2)  (h_i^2+r^2)^{\frac{\alpha_i}{2}}}{\rho_i \eta_i })\Big\}\nonumber, \\
	&\overset{(a)}{=}\mathbb{E}_{\mathcal{S},r,i|r,I|r,i,\mathcal{S}}  \Big\{\frac{\Gamma(m_i,m_i (I+\sigma^2) X)}{\Gamma(m_i)} \Bigg|_{X= \frac{ \gamma (h_i^2+r^2)^{\frac{\alpha_i}{2}}}{\rho_i \eta_i } }\Big\}\nonumber, \\
	&\overset{(b)}{=}\mathbb{E}_{\mathcal{S},r,i|r,I|r,i,\mathcal{S}}  \Big\{\exp\left(-m_i(I+\sigma^2) X\right) \sum_{k=0}^{m_i-1} \frac{m_i^k (I+\sigma^2)^k X^k}{k!} \Big\}\nonumber, \\
	&=\mathbb{E}_{\mathcal{S},r,i|r
	}\Bigg\{ \mathbb{E}_{I|r,i,\mathcal{S}}\Big( \exp\left(-m_i X (I+\sigma^2)\right) (I+\sigma^2)^k \Big)\sum_{k=0}^{m_i-1} \frac{m_i^k  X^k}{k!} \Bigg\}\nonumber, \\
	&=\mathbb{E}_{\mathcal{S},r,i|r}\Bigg\{ \sum_{k=0}^{m_i-1}(-1)^k  \frac{{s_r}^k}{k!}  \frac{d^k}{{ds_r}^k}\mathcal{L}_{I+\sigma^2}(s_r,r,i, \mathcal{S})\Bigg\}\nonumber, \\
	&\overset{(c)}{=}\mathbb{E}_{\mathcal{S}}\Bigg\{ \int_{0}^{\infty}\sum_{i \in \mathcal{A}}^{} \sum_{k=0}^{m_i-1}(-1)^k  \frac{{s_r}^k}{k!}  \frac{d^k}{{ds_r}^k}\mathcal{L}_{I+\sigma^2}(s_r,r,i, \mathcal{S}) A_i(r,i,\mathcal{S}) f_{R_i}(r) \nonumber \\& + \sum_{i \in \mathcal{B}}^{} \sum_{k=0}^{m_i-1}(-1)^k  \frac{{s_r}^k}{k!}  \frac{d^k}{{ds_r}^k}\mathcal{L}^c_{I+\sigma^2}(s_r,r,i) A_i^c(r,i) \mathcal{P}_i(r) f_{r_{cu|\mathcal{S}}}(r) dr\Bigg\}.
\end{align}
 In (a), since $G_i$ follows a Gamma distribution, we use the CDF of a Gamma distribution, (b) comes form the power series expansion of the lower incomplete gamma function, and in (c), we average over the distance to the serving node and the channel link type (LOS,NLOS). 
\bibliographystyle{IEEEtran}
\bibliography{hokie-HD}

\end{document}